## ToC Graphic

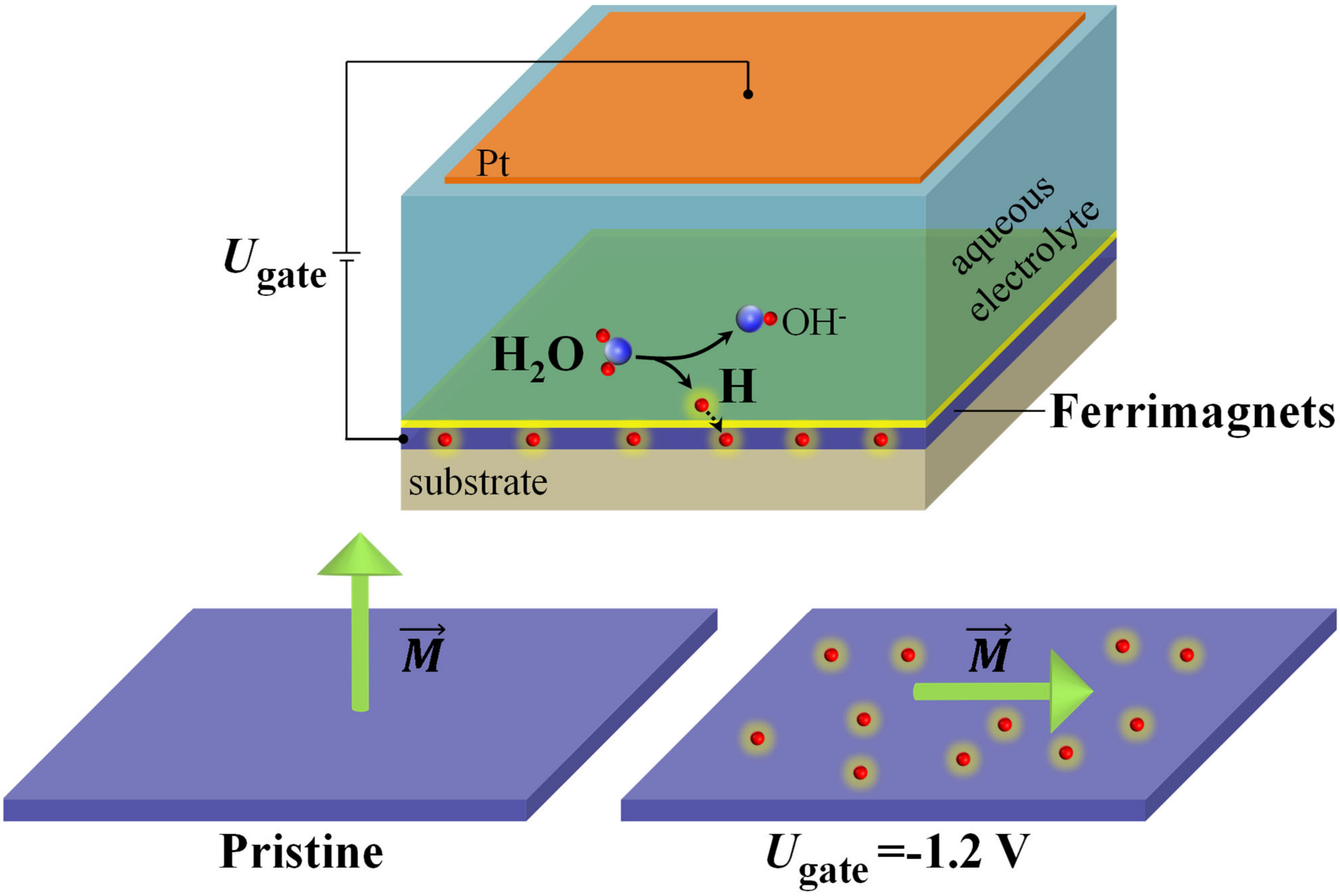

Pt
$U_{gate}$
aqueous electrolyte
$H_2O$
$OH^-$
H
Ferrimagnets
substrate
$\vec{M}$
$\vec{M}$
Pristine
$U_{gate}$ =-1.2 V

# Voltage-gated 90° switching of bulk perpendicular magnetic anisotropy in ferrimagnets

Zhengyu Xiao[1,2,3♪], Ruiwen Xie[4♪], Fernando Maccari[4], Philipp Klaßen[5], Benedikt Eggert[5], Di Wang[2,6], Yuting Dai[2], Raquel Lizárraga[7,8], Johanna Lill[5], Tom Helbig[5], Heiko Wende[5], Kurt Kummer[9], Katharina Ollefs[5], Konstantin P Skokov[4], Hongbin Zhang[4], Zhiyong Quan[3,10], Xiaohong Xu[3,10], Robert Kruk[2], Horst Hahn[2,11], Oliver Gutfleisch[4], Xinglong Ye[4,1,12*]

[1]TUD-KIT Joint Research Laboratory Nanomaterials, Technische Universität Darmstadt, 64287 Darmstadt, Germany
[2]Institute of Nanotechnology, Karlsruhe Institute of Technology, 76344 Eggenstein-Leopoldshafen, Germany
[3]Key Laboratory of Magnetic Molecules and Magnetic Information Materials of Ministry of Education, School of Chemistry and Materials Science, Shanxi Normal University, 030032 Taiyuan, China
[4]Institute of Materials Science, Technische Universität Darmstadt, 64287 Darmstadt, Germany
[5]Faculty of Physics and Center for Nanointegration Duisburg-Essen (CENIDE), University of Duisburg-Essen,
47057 Duisburg, Germany
[6]Karlsruhe Nano Micro Facility, Karlsruhe Institute of Technology (KIT), 76131 Karlsruhe, Germany
[7]Applied Materials Physics, Department of Materials Science and Engineering, Royal Institute of Technology, Stockholm SE-100 44, Sweden
[8]Wallenberg Initiative Materials Science for Sustainability, Royal Institute of Technology, Stockholm SE-100 44, Sweden
[9]European Synchrotron Radiation Facility, BP 220, F-38043, Grenoble Cedex, France
[10]Collaborative Innovation Center for Shanxi Advanced Permanent Magnetic Materials and Technology, Research Institute of Materials Science, Shanxi Normal University, 030032 Taiyuan, China
[11]School of Sustainable Chemical, Biological and Materials Engineering, University of Oklahoma, Norman OK 73019, United States
[12]Present address: School of Physics, Shandong University, 250100 Jinan, China

♪ These authors contribute equally to this work.
* xinglong.ye@tu-darmstadt.de

## ABSTRACT

Unravelling the mechanism behind bulk perpendicular magnetic anisotropy (PMA) in amorphous rare earth-transition metal films has proven challenging. This is largely due to the inherent complexity of amorphous structure and the entangled potential origins arising from microstructure and atomic structure factors. Here, we present an approach wherein the magneto-electric effect is harnessed to induce 90° switching of bulk PMA in Tb-Co films to in-plane directions by applying voltages of only -1.2 V. This manipulation is achieved by voltage-driven insertion of hydrogen atoms into interstitial sites between Tb and Co atoms, which serves as a perturbation to local atomic structure. Using angle-dependent X-ray magnetic circular dichroism, we find that the anisotropy switching originates from the distortion of crystal field around Tb which reorients the alignment of Tb orbital moments. Initially aligned along Tb-Co bonding directions, the easy magnetization axis undergoes reorientation and switches by 90°, as substantiated by ab-initio calculations. Our study not only concludes the atomic origin of Tb-Co atom bonding configuration in shaping bulk PMA, but also establishes the groundwork for electrically programmable ferrimagnetic spintronics, such as controlling domain wall motion and programming artificial spin textures.

## INTRODUCTION

Ferrimagnets, which combine the advantages of both ferromagnets and antiferromagnets — namely controllable magnetization and antiferromagnetic-like fast dynamics — have driven the rapid emergence of ferrimagnetic spintronics.[1] Among ferrimagnetic materials, amorphous rare earth-transition metal (RE-TM) ferrimagnetic films exhibit extraordinary properties including ultrafast domain wall movement,[2,3] field-free magnetization reversal,[4] bulk Dzyaloshinskii–Moriya interaction,[5] nanoscale skyrmions,[6] flipping handedness of magnons[7] and femtosecond optical switching.[8-10] These properties exploit the unique advantage of bulk perpendicular magnetic anisotropy (PMA) in amorphous ferrimagnetic films, i.e. a preferred magnetization orientation along out-of-plane directions in films even with large thickness (e.g. 500 nm).[11,12] While the discovery of PMA dates back to 1973, its manifestation in amorphous materials raises fundamental questions about its origin. Intrinsically, the local magnetic anisotropy originates from the interaction between aspherically-distributed 4*f* electrons of RE ions, characterized by strong spin-orbital coupling, and the surrounding aspherical crystal field. In crystalline materials, the alignment of this local anisotropy with long-range crystal symmetries gives rise to macroscopic magnetic anisotropy, as seen in high-performance permanent magnets such as $SmCo_5$ and $Nd_2Fe_{14}B$.[13,14] However, in amorphous alloys, which lack crystal symmetry, the local single-ion anisotropy of RE ions exhibits variations in both directions and magnitudes from site to site, effectively canceling out at macroscopic scales.[15]

Understanding the origin of bulk PMA has been hindered by the inherent difficulty of determining precise atomic positions and structure within amorphous structure.[16,17] Proposed models to explain bulk PMA encompass both extrinsic microstructure and intrinsic atomic structure factors, such as columnar morphology,[18] composition inhomogeneity,[19] magneto-elastic effect,[20] pair-order anisotropy,[21,22] dipolar effect,[23] and bond-orientation anisotropy.[24] However, the exact mechanism governing PMA remains elusive. This is largely because that conventional approaches to modulating bulk PMA energy, such as altering magnetron sputtering parameters, often results into the concomitant changes of multiple microstructural and atomic-scale features. This complicates the interpretation of experimental results. Disentangling effectively these changes and separating their effects on bulk PMA can be the key to resolving the mechanism of bulk PMA.

Recent studies have demonstrated that magnetic properties of materials, including magnetic anisotropy, coercivity and Curie temperatures, can be controlled by applying small voltages, as exemplified in many ferro-/ferri-magnetic materials.[25-34] Unlike sputtering techniques, the magneto-electric approach modifies magnetism through mechanisms of electron doping or migration of ions (magneto-ionics), which usually does not change the microstructure of materials. In particular, we showed that the magnetic properties of ferromagnetic intermetallics even in micrometer scale can be substantially modulated through electrochemically-controlled insertion and extraction of hydrogen atoms in metal structure.[35,36] Unlike electrons and ions, hydrogen atoms are electrically neutral, and, therefore, their diffusion into the metal structure is not restricted by electric-field screening. This enables the control of bulk magnetic properties with small voltages. Moreover, hydrogen insertion into interstitial sites distorts atomic structure while preserving the microstructure of materials, such as grain sizes and phase distributions.[32,33] In the case of amorphous ferrimagnets, we expect that hydrogen insertion into interstitial sites will not change the microstructural features such as column texture, roughness, geometry, spatial distribution of elements and thickness at micro/nanometer scale. This capability thus enables the disentanglement of the effects of the microstructure on bulk PMA, providing insights into the atomic origin of bulk PMA. In this work, we employed the newly-developed approach of voltage-driven hydrogen insertion/extraction approach to modify bulk PMA in amorphous ferrimagnets. We show that by applying voltages as small as -1.2 V, bulk PMA of amorphous ferrimagnetic TbCo films can be switched completely by 90° to in-plane directions. This yields a voltage-controlled huge modification of coercivity by 2.0 T at room temperature, the largest ever achieved by magneto-electric approaches. Through angle-dependent X-ray magnetic circular dichroism (XMCD), we find that upon hydrogen insertion, Tb orbital moments, which determine the single-ion magnetic anisotropy, reorients from out-of-plane direction to in-plane directions due to distortion of crystal field. As verified by ab-initio calculations, the easy axis of magnetization initially aligns along Tb-Co bonding directions, but reorients and switches their directions due to the presence of interstitial hydrogen atoms. Our findings conclude the long-standing question on the atomic origin of bulk PMA in amorphous ferrimagnets. Moreover, the ability to manipulate PMA with small voltages of about 1 V holds great promise for significantly advancing electrically-programmable ferrimagnetic spintronics.

## RESULTS AND DISCUSSION

We used amorphous TbCo thin films as our model material, renowned for its large bulk PMA.[37,38] Thin films were deposited onto thermally-oxidized Si substrates by magnetron co-sputtering of Tb and Co elements (purity 99.9%) at room temperature, with thicknesses of 20 nm, 50 nm and 500 nm. Compositions were adjusted by varying sputtering powers which leads to magnetization compensation temperatures ($T_M$) from 200 K to above room temperature. We further deposited 5 nm-thick Pd layers onto TbCo films to avoid sample oxidation, which also facilitates the passage of hydrogen atoms. To control insertion of hydrogen atoms in TbCo films, we utilized an electrochemical setup with three electrodes (Fig. 1A), i.e. the working electrode (WE, Pd/TbCo/Si), the counter electrode (CE, Pt foil), and the reference electrode (Hg/HgO electrode, not shown for clarity). The electrolyte is 1 M KOH aqueous electrolyte (shaded by light green color). When a gate voltage (-1.2 V) was applied between the working electrode and the counter electrode across the electrolyte, water molecules were reduced into hydrogen atoms that initially adsorb onto Pd surface. Driven by gradient of concentration, they diffuse then into the Pd layer and subsequently into the TbCo layer. Notably, the mixing enthalpy of hydrogen atoms and TbCo is more negative than that of hydrogen and Pd.[39,40] Consequently, hydrogen atoms exhibit unidirectional diffusion from Pd to TbCo layers, ensuring the non-volatile and irreversible retention of voltage-driven hydrogen insertion within TbCo layers. This irreversibility facilitates subsequent characterization of magnetic properties, electron transport and magnetic moments under high vacuum and at low temperatures.

As-grown TbCo thin films exhibits typical PMA at various temperatures evidenced by the rectangular out-of-plane hysteresis loops probed by anomalous Hall effect (AHE) (Fig. 1B). At 300 K the polarity of AHE resistance remains negative under positive magnetic field, indicating the compensation temperature $T_M$ above 300 K (with nominal composition $Tb_{45}Co_{55}$).[41] In comparison, upon applying -1.2 V, the $T_M$ decreased substantially to about 210 K, signified by the inversion of AHE polarity (Fig. 1C). As a result, the coercivity at room temperature changes from 2.5 T to 0.5 T, representing an unprecedented voltage-induced coercivity manipulation by 2 T. Most importantly, we observed that after applying -1.2 V the AHE hysteresis loops exhibited the hard axis behavior with tilted shapes under out-of-plane magnetic fields, indicating the 90° switching of magnetic anisotropy from PMA to in-plane directions. Likewise, we have observed the similar voltage-driven anisotropy switching for other TbCo thin films with initial $T_M$ of 200 K (Fig. S1, S2) and 295 K (Fig. S3).

Magnetometry measurements validate the voltage-driven anisotropy switching through directly measuring magnetization behaviors along both out-of-plane and in-plane directions on TbCo thin films (initial $T_M$ of 200 K). Fig. 2A, 2C show the magnetic hysteresis loops of TbCo samples before and after applying -1.2 V at room temperature (for measurements at low temperature please see Fig. S4). By comparing hysteresis loops along two directions, it is clear that the as-grown sample exhibits the typical PMA, in agreement with AHE measurements (Fig. S1, S2). In contrary, after applying −1.2 V, the magnetic hysteresis loops interchanged their shapes along two directions, with the in-plane loop now displaying the rectangular shape (Fig. 2C). This verifies the reorientation of the magnetic anisotropy from the perpendicular to the in-plane direction. With the measured values of 5 T for anisotropy field ($H_a$), the anisotropy energy was calculated to be 0.5 MJ/m$^3$ at 100 K using the relation $K_U = 1/2H_aM_S + 1/2\mu_0M_S^2$,[42] indicating the giant magnetoelectric efficiency. Note that the magnetization drop at low magnetic fields for all samples can be attributed to the Tb-rich layer near the substrate (Fig. S5, S6).

Another intriguing feature induced by voltage-driven anisotropy switching is the atypical magnetization reversal process observed by magneto-optical Kerr effect (MOKE) microscopy. For as-grown samples, the reversed magnetic domains nucleate at multiple positions from the fully-magnetized state and then propagate into neighboring regions (Fig. 2A,2B). These reversed domains have fractal, irregular boundaries and creep fast into non-reversed regions under constant magnetic fields, a phenomenon known as magnetic aftereffect.[43] In contrast, after voltage-driven hydrogen insertion, magnetization reversal proceeds predominantly through point-by-point reversals of interspersed magnetic domains (Fig. 2C,2D). Those domains exhibit no further growth, and under increasing fields more individual reversed domains occur until the full reversal of magnetization.

These comprehensive AHE, magnetometry and MOKE measurements thus clearly demonstrate the ability to manipulate the magnetic anisotropy of ferrimagnetic thin films from perpendicular to in-plane directions by applying small voltages. This voltage-driven anisotropy switching, therefore, presents a unique opportunity to probe the origin of PMA by elucidating the structural changes induced by hydrogen insertion. Our investigation commenced with the microstructure characterization. Transmission electron microscopy (TEM) reveals that after hydrogen insertion, TbCo thin films maintain their homogenous morphology without forming cracks or substructures (Fig. S5). Energy dispersive X-ray spectroscopy (EDS) mapping shows inhomogeneous distributions of Tb and Co atoms, typical of amorphous thin films sputtered at room temperature,[5,44] but no discernable differences were detected before and after hydrogen insertion (Fig. S6, S7). Hydrogen distribution can be inhomogeneous with potentially higher concentration near the Pd/TbCo and TbCo/substrate interfaces. This is due to non-uniform distribution of Tb elements which tend to segregate at these interfaces (Fig. S6, S7). However, hydrogen segregation is confined to about 2 nanometers near the interfaces and its effect on magnetic anisotropy is likely insignificant, as the phenomena of voltage-driven anisotropy switching has been observed even in thick films (500 nm). Selected area electron diffraction (SAED, Fig. S5) and macroscopic X-ray diffraction (XRD, Fig. S8) confirm the maintenance of amorphous structure after voltage-driven hydrogen insertion, with broad, diffusive diffraction peaks/halos. These results align with our anticipation that hydrogen insertion into interstitial sites between Tb and Co atoms will not change the microstructural features. Moreover, the magnetic anisotropy of RE-TM alloys, as mentioned earlier, originates from the interaction between strongly spin-orbit-coupled, aspherically distributed 4f electrons in rare earth elements and the surrounding crystal field created by neighboring atoms. It is considered that the 4*f* electrons are deeply localized and shielded from the external environment, making them less susceptible to external environmental changes. Given this, the anisotropy switching from PMA to in-plane directions is more likely attributed to the distortion of the crystal field surrounding the Tb atoms.

Given the inherent difficulty of directly resolving the atomic structure and its distortion in amorphous materials, we probed the electronic structure and magnetic moments of Tb and Co through element-specific X-ray absorption (XAS) and XMCD measurements.[45] These measurements were conducted along both grazing and normal directions (15° and 70° with respect to surface normal) on TbCo films (initial $T_M$ of 295 K, Fig. S4). Fig. 3A shows that the XAS spectra measured at Co $L_{2,3}$ edges and Tb $M_{4,5}$ possess no significant changes in line shapes and peak positions before and after hydrogen insertion, indicating the unchanged metallic character for Co[46,47] and the $Tb^{3+}$ states.[48] However, the appearance of the spectral fine structures at Co $L_3$ white line and a small shift towards lower energies (see inset of Fig. 3A) after hydrogen insertion suggests the presence of interstitial hydrogen near Co atoms.

We resolved the spin and orbital moments of Tb and Co elements by applying magneto-optical sum rules to the XMCD spectra (Fig. 3B,C).[49,50] For the Tb $M_{4,5}$ edges, we used ab-initio calculations to calculate the contribution from the magnetic dipole operator $\langle T_z \rangle$, which can be significant in RE element in contrast to transition metals. The values obtained are $\langle T_z \rangle$ = 0.20 (0) for 15° and 0 (0.21) for 70° for as-grown (hydrogen-charged) samples, which agree well the values calculated in the literature.[51] We refer the reader to the supplemental materials for detailed discussions of the obtained $\langle T_z \rangle$ term and sum rule analysis (Fig. S9-11). As outlined in Fig. 3C and Table S1, the total magnetic moment of Co atom in the as-grown samples is -2.5 $\mu_B$, consistent with XMCD results for other amorphous RE-Co compounds.[51] The large magnitude of the Co moment might be associated with the increased bond distances in amorphous alloys as compared to *hcp*-Co. For Tb, the total magnetic moment and the orbital moment are 8.9 $\mu_B$ and 3.7 $\mu_B$, matching the expected values for the $4f^{8}$ configuration and consistent with previously reported data.[51] Upon hydrogen insertion, the magnetic moments of both Tb and Co decreased substantially by 35% and 50%, respectively (Fig. 3C), which are associated with the observed decrease of $T_M$. Of particular interest is the angle-dependence of magnetic moments of Tb before and after hydrogen insertion, which are directly related to magnetic anisotropy according to the RE single-ion anisotropy model. For the as-grown samples, we observed larger magnetic moments and orbital moments of Tb in perpendicular directions compared to grazing incidence directions. This phenomenon, i.e. orbital moments of RE atoms are larger along perpendicular directions than grazing directions, has also been observed in amorphous NdCo films with large PMA.[52] However, after hydrogen insertion, the opposite was observed, i.e., the Tb orbital moments become larger along the grazing angle than along perpendicular directions. This indicates the reorientation of local magnetic anisotropy at atomic scale and the change of the crystal field around Tb. It is worthwhile to note that, due to the limited probing depth of XMCD measurements in total electron yield mode (< 10 nm), the magnetic moments obtained from XMCD analysis may differ from those of the entire films, as Tb elements tend to segregate near the surface. However, since bulk PMA and the voltage-driven anisotropy switching have been observed throughout the entire films, the trend in how magnetic moments change along two directions before and after hydrogen insertion is expected to remain consistent.

We noted that although the magnetic moments of Tb and Co at 5 K, as obtained from XMCD, decrease substantially after hydrogen insertion, the overall saturation magnetization measured at 300 K via magnetometry remains almost unchanged (Fig. 2A, 2C). This discrepancy can be attributed to three factors. Firstly, XMCD probes element-specific magnetic moment of Tb and Co atoms, whereas magnetometry measures the net magnetization associated with the difference between Tb and Co moments. Consequently, when both Tb and Co moments decrease, their net subtraction may exhibit less pronounced decrease. Secondly, due to antiferromagnetic ordering of Tb and Co moments, the net saturation magnetization of TbCo films depends strongly on the temperature relative to their compensation temperatures. The $T_M$ for the pristine samples and those after hydrogen insertion, are 198 K and 130 K, respectively (Fig. 2, Fig. S2). When measured at 300 K, the pristine samples are closer to their compensation temperatures than those after hydrogen insertion. As a result, the net magnetization of the pristine samples decreases more significantly, making their values comparable to those after hydrogen insertion. Thirdly, XMCD primarily probes near-surface region with a typical penetration depth of less than 10 nm, whereas magnetometry measures the entire films. In our cases, Tb elements tend to segregate at surfaces, leading to the differences in the composition of the near-surface region compared to the overall film. This distinction in measurement techniques further complicates the direct comparison between XMCD and magnetometry results. Given these factors, we primarily use magnetometry to verify the voltage-driven anisotropy switching since it can measure magnetization from both perpendicular and in-plane directions. In complementary, we used XMCD to probe the spin and orbital moment of Tb and Co atoms, providing insights into the changes in their electronic environment induced by interstitial hydrogen atoms. We also notice that the obtained anisotropy in magnetic moments of Tb and Co, measured along two different angles by XMCD, are relatively larger than that obtained from magnetometry measurements. In the latter case, saturation magnetization along hard axis can approach that along easy axis especially at 300 K (Fig. 2A, 2G, S4). This difference can be again caused by the distinction between XMCD and magnetometry techniques, which measure near-surface regions and the overall films, as well as the strong dependence of magnetic moments on temperatures.

To elucidate the distortion of the crystal field and the corresponding anisotropy switching at atomic scale induced by hydrogen insertion, we acquired the three-dimensional mapping of magnetic anisotropy in amorphous TbCo and TbCo-H structures by ab initio calculations. The amorphous TbCo structure was simulated by constructing a 150-atom supercell with 25 Tb and 125 Co atoms using stochastic quenching (Fig. 4A, Fig. S12).[53,54] The single-ion anisotropy of each Tb atom was evaluated by solving its atomic Hamiltonian, including coulomb interactions ($\widehat{H}_U$), spin-orbit coupling (SOC), crystal field Hamiltonian ($\widehat{H}_{CF}$) and the exchange coupling term resulted from the magnetization of Co atoms ($\widehat{H}_{ex}$)

$$\hat{H}_{at} = \hat{H}_U + \lambda \sum_i \hat{s}_i \hat{l}_i + \hat{H}_{CF} + \hat{H}_{ex}.$$

In particular, the crystal field parameter $\hat{H}_{CF} = \sum_{k=0}^{k_{max}} \sum_{q=-k}^{k} B_q^{(k)} \hat{C}_q^{(k)}$ was obtained by transforming the Tb 4*f* band states calculated by the density functional theory (DFT) to Wannier functions, which were expanded with crystal field parameters (CFP) $B_q^{(k)}$ in a series of spherical tensor operators $\hat{C}_q^{(k)}$.[55] Subsequently, we evaluated the magnetic anisotropy ($E_{\text{aniso}} = E_{\text{eigen}} - E_{\text{eigen}}^{\text{min}}$) by calculating the eigenvalue ($E_{\text{eigen}}$) of $\hat{H}_{\text{at}}$ with magnetization directions defined by polar angle $\theta$ ([0, π]) and azimuth angle $\varphi$ ([−π, π]).

Analyzing the correlations between $E_{\text{aniso}}$ and Tb-Tb/Tb-Co bonding directions, we found that the easy magnetization directions are statistically aligned along the Tb-Co bonding directions (Fig. S13). This feature persists across various compositions, for instance $TbCo_4$ (Fig. S13). Fig. 4B,D show the contour plots of $E_{\text{aniso}}$ against θ and $\varphi$ for two Tb atoms (indexed as Tb12 and Tb19) among 25 Tb atoms, along with the angular projections of Tb-Co and Tb-Tb bonding vectors. It is evident that the low/high $E_{\text{aniso}}$ zones, i.e., the easy/hard magnetization directions, significantly overlapped with the Tb-Co/Tb-Tb bonding directions. This important finding pinpoints the critical role of the Tb-Co bonding direction in shaping PMA. Previously, Harris et al. observed in TbFe thin films that the Fourier-transformed density of TbFe nearest neighbor pairs along the out-of-plane direction is slightly larger (1-4%) than along in-plane directions, thereby leading to the phenomenological model of pair-order anisotropy.[21,22] However, the doubts remain about how the PMA arise at atomic scale and whether such minor anisotropy can induce such high PMA.[56] Our finding, for the first time, reveals the preference of magnetization along Tb-Co directions at atomic scale. Most compellingly, upon hydrogen insertion as a perturbation, the initially high $E_{\text{aniso}}$ zone (Tb-Tb directions) decreased to values comparable to those along Tb-Co directions, thereby substantially weakening the magnetic anisotropy (Fig. 4C). Remarkably, for the atom configurations of Tb12, the crystal field parameter $B_2^0$ changed its sign from positive to negative values upon hydrogen insertion (Fig. 4E), which causes a 90° reorientation of the magnetization axis (Fig. S14,15, Table S2).[57,58] This result reproduced the experimentally observed 90° anisotropy switching, indicating the representative atom configuration for large PMA. These theoretical calculations, using interstitial hydrogen atoms as a perturbation to atomic structure, thus clarify the distortion of the crystal field around Tb in switching PMA, as experimentally observed.

## CONCLUSION

In summary, by combining experimental and theoretical approaches we explored the atomic origin of bulk PMA in amorphous ferrimagnetic films. To disentangle the effect of microstructure, we employed the recently-developed magneto-electric approach of voltage-driven insertion of hydrogen atoms into interstitial sites, which acts as a perturbation to the local atomic structure. Remarkably, we show that by applying only -1.2 V the bulk PMA can be switched by 90° to in-plane directions in TbCo films. This enables the unprecedented change of coercivity by 2.0 T at room temperature and the evolution of interspersed magnetic domain structures, observed for the first time in ferrimagnets. Analyzing the anisotropy-switching process, angle-dependent XMCD reveals the loss of preferential orientation of Tb orbital moments, which determines the single-ion anisotropy of Tb atoms. As verified by ab-initio calculations, the easy magnetization axis originally aligns along Tb-Co bonding directions, but reorients and switches by 90° due to the distorted crystal field. Our study disentangles the effect of microstructure and atomic structure on bulk PMA, identifying the decisive role of the Tb-Co atom configuration in shaping PMA. An exceptionally large PMA may thus be anticipated by depositing sequential layers of RE and TM elements, magnifying the population of Tb-Co bondings along perpendicular directions. Moreover, our work opens up a previously inaccessible route to manipulate local PMA in ferrimagnetic films by applying small voltages. This lays the groundwork for broad applications in the rapidly emerging field of ferrimagnetic spintronics,[1-10] such as controlling domain-wall movement, skyrmion sizes, and magnon propagation. The ability to engineer local magnetic anisotropy with ultralow voltages also enables the development of electrically-programmable artificial spin textures in ferrimagnets, such as those with alternating in-plane and out-of-plane magnetization as building blocks.[59,60]

## EXPERIMENTAL METHODS

### Growth of thin films

Amorphous TbCo thin films were deposited by DC magnetron co-sputtering of Tb and Co targets at room temperature. Considering that the PMA is bulk magnetic anisotropy, we deposited thin films directly onto thermally-oxidized Si substrate without using buffer layers. The background pressure of the sputtering chamber was $<9\times10^{-8}$ mbar, and the Ar pressure was 3 mbar during deposition. The TbCo compositions were adjusted by varying the Tb sputtering power to tune the Tb deposition rate. The layer thicknesses and nominal alloy atomic compositions were determined by EDS and TEM. A 5 nm Pd layer was deposited onto TbCo layer as a capping layer, which prevented the oxidation of Tb-Co layer and allowed the transport of hydrogen atoms into TbCo layer during voltage-driven hydrogen insertion. For XMCD measurements, we used a Pd capping layer with a smaller thickness of 2 nm to increase the electron yield from TbCo layers.

### Magnetometry, AHE and MOKE characterizations

Magnetometry measurements were performed in a superconducting quantum interference device (SQUID, MPMS3) at different temperatures with magnetic fields applied both along out-of-plane and in-plane directions. The magnetic hysteresis loops of TbCo layers were obtained by subtracting the linear diamagnetic contribution of Si substrate from the measured magnetization. The anomalous Hall effect (AHE) measurements were performed in PPMS at different temperatures utilizing the van der Pauw method. The samples were cut into square shapes with a size of approximately 4 mm $\times$ 4 mm. The AHE mostly arises from by the Co sublattice, and thus its polarity allowed us to determine the magnetically dominant sublattice. The evolution of magnetic domain structure during magnetization reversal process was characterized by magneto-optical Kerr effect (MOKE) microscope (Zeiss Axio Imager, D2m evico magnetics GmbH, wavelength 660 nm) under magnetic fields using both polar and longitudinal modes. To enhance the image contrast, the non-magnetic background image was subtracted from the collected average image using KerrLab software.

### Electrochemical charging of hydrogen atoms

Voltage-driven hydrogen insertion into TbCo films were carried out under potentiostatic control in a three-electrode electrochemical system (Autolab PGSTAT 302N). The working, counter, and reference electrodes were Tb-Co thin films, Pt wires and a pseudo Ag/AgCl electrode, respectively, within the home-made electrochemical cell in PPMS. To charge the thin film, we used the potential of -1.2 V for a short duration of about 2 minutes. The potential of the peuso Ag/AgCl electrode is 0.300±0.002 V more positive than the standard Hg/HgO (1M KOH) electrode, and all potentials in the paper were with reference to Hg/HgO (1M KOH) electrode. The electrolyte was an aqueous electrolyte of 1 M KOH prepared from ultrapure water with a resistivity of ~ 18.2 MΩ cm.

**TEM characterization**

The microstructure of amorphous thin films were characterization by TEM and selected area diffraction (TEM, FEI Titan 80-300) equipped with EDS. Preparation of TEM samples followed the procedure of cutting, lifting and milling using FIB/SEM dual beam system (FEI Strata 400 and Zeiss Auriga 60, KIT). Radial distribution functions of elements were obtained by Fourier transformation of the selected area diffraction pattern collected at least at four different positions.

**XMCD measurements**

X-ray absorption spectroscopy (XAS) and X-ray magnetic circular dichroism (XMCD) experiments were performed at the ID32 beamline of the European Synchrotron (ESRF, Grenoble, France) by measuring absorption spectra at the Co $L_{2,3}$ and Tb $M_{4,5}$ edges. All spectra were recorded using total electron yield detection mode. The isotropic XAS spectra were approximated by averaging XAS spectra measured with right and left circularly polarized X-rays and under constant magnetic field of 8 T, whereas XMCD spectra were obtained as their difference. All spectra for each sample before and after voltage-driven hydrogen insertion were measured at two different sample orientations, i.e. in normal (15°) and grazing angles (70°) with respect to the normal of sample surface. The expectation values for the spin and orbital moments of Co and Tb were calculated using the following dichroism sum rules.

$$\langle S_z^{Co} \rangle = (10-n)\frac{\int_{L_3} d\omega(\mu^+ - \mu^-) - 2\int_{L_2} d\omega(\mu^+ - \mu^-)}{\int_{L_{2,3}} d\omega(\mu^+ + \mu^-)} - \frac{7}{2}\langle T_z \rangle$$

$$\langle L_z^{Co} \rangle = (10-n)\frac{4}{3}\frac{\int_{L_{2,3}} d\omega(\mu^+ - \mu^-)}{\int_{L_{2,3}} d\omega(\mu^+ + \mu^-)}$$

$$\langle S_z^{Tb} \rangle = (14-n)\frac{\int_{M_5} d\omega(\mu^+ - \mu^-) - \frac{3}{2}\int_{M_4} d\omega(\mu^+ - \mu^-)}{\int_{M_{4,5}} d\omega(\mu^+ + \mu^-)} - 3\langle T_z \rangle$$

$$\langle L_z^{Tb} \rangle = (14-n)2\frac{\int_{M_{4,5}} d\omega(\mu^+ - \mu^-)}{\int_{M_{4,5}} d\omega(\mu^+ + \mu^-)}$$

Here, $n$ is the number of occupied states in the outer shell and $\langle T_z \rangle$ are the expectation values for the magnetic dipole operator. For the extraction of the Co moments, we assumed $n = 2.49$[61] and for simplicity $\langle T_z \rangle = 0$. For Tb we assumed $n = 8$ ($Tb^{3+}$). To derive the respective moments, we used the known relations $\mu_S = -2\langle S_z \rangle\mu_B$ and $\mu_L = -\langle L_z \rangle\mu_B$. The expectation values $\langle T_z \rangle$ for the Tb atoms were calculated using ab initio calculations and averaged over all Tb atoms, and the calculated results are shown in Fig. S9. The dependence of the magnetic moments on the $\langle T_z \rangle$ for the range of the calculated values is presented in Fig. S10. It is known that $\langle S_z \rangle$ and $\langle T_z \rangle$ are antiparallel for $Tb^{3+}$ ions, and consequently only positive values were considered when $\langle S_z \rangle$ is negative. To account for the transitions into the continuum, we have deployed step functions and subtracted it from the measured X-ray absorption intensity to form the integrant $(\mu^+ + \mu^-)$ as the denominator in the above formulated sum rules. The used step functions are shown in Fig. S11.

**Ab initio calculations**

The stochastic quenching procedure for the construction of the 150-atom cell is as follows: an initial configuration of 150 atoms was generated by randomly distributing them in a cubic box with the constraint that the closest atom pair was 2.0 Å. The atomic coordinates were then relaxed using the Vienna ab-initio simulation package (VASP) until the forces on every atom were smaller than 0.05 eV/Å.62 The lattice parameter of the cubic cell was determined to be around 12.0 Å. In the VASP calculation, the $4f$ electrons of Tb were treated as core states. The calculation was performed at the Γ-k point with an energy cutoff of 350 eV. Two structures were constructed to simulate Tb-Co amorphous systems with different Co/Tb ratios, denoted as Tb25Co125 and Tb32Co118, respectively (see Fig. S12 (A) and (C)). To simulate the hydrogen-charged amorphous Tb-Co system, we added one extra hydrogen atom into each Tb-centred cluster in the pristine structures, denoted as Tb25H25Co125 and Tb32H32Co118, respectively (see Fig. S11 (B) and (D)).

In terms of the evaluation of crystal field parameters (CFPs), we followed the method proposed by Novák et al. that uses Wannier functions to construct the crystal field Hamiltonian $\hat{H}_{CF} = \sum_{k=0}^{k_{max}} \sum_{q=-k}^{k} B_q^{(k)} \hat{C}_q^{(k)}$, where $\hat{C}_q^{(k)}$ is a spherical tensor of rank k acting on electrons in the 4f shell. $B_q^{(k)}$ are CFPs. The calculation of CFPs consists of four steps:

(1) The self-consistent band calculation by treating $4f$ electrons of Tb as core states was performed to determine the single-particle potential of the Tb ion.

(2) The $4f$ states as well as Co $3d$ states (for hydrogen-charged systems the H $1s$ states as well) were treated as valence states in a non-self-consistent calculation while all other states were moved away using the orbital shift operator. In addition, a correction $\Delta$ that amounts to the downward shift of Co $3d$ (H $1s$) level was introduced to approximate the actual charge transfer energy by modifying the difference $\epsilon_f - \epsilon_d$.

(3) The Tb $4f$ states were transformed to Wannier basis using the Wien2wannier[63] and Wannier90[64] packages. Wannier90 provides the atom-cantered $7 \times 7$ matrix $\hat{H}_{4f}$, which is equivalent to $E_{avg}\hat{I} + \sum_{k,q} B_q^{(k)} \hat{C}_q^{(k)}$ with $E_{avg} = Tr(\hat{H}_{4f}/7)$. The traceless part is then the desired crystal field Hamiltonian $\hat{H}_{CF}$.

(4) To get the standard form of CFPs, $\hat{H}_{4f}$ was transformed into the basis of spherical harmonics and expanded as a 49-dimension vector in the basis of spherical tensor operators.

For the evaluation of CFPs, the optimized pristine and hydrogen-charged TbCo structures using VASP were employed and the Tb atoms were assumed to possess $Tb^{3+}$ state. The steps (1) and (2) were carried out using the WIEN2k package with implemented augmented plane waves + local orbital method. [65] The generalized-gradient approximation (GGA) was used as the exchange-correlational functional. We set $RK_{max} = 7.0$ for the Γ $k$ point calculation. The atomic radii of Tb and Co were 2.23 and 1.98, respectively. In step (2), the hybridization parameters $\Delta$ corresponding to Co $3d$ and hydrogen $1s$ states were taken as -0.6 and -1.4 Ry, respectively, to assure that the Wannier functions of Tb $4f$ orbitals are located at the Tb atomic positions with very limited spread (smaller than 0.5 $Å^2$ in this case).

In order to evaluate the single-ion anisotropies (SIAs) of Tb, we constructed the atomic Hamiltonian of Tb by including the Coulomb interactions ($\hat{H}_U$), the spin-orbit coupling (SOC), the crystal field

Hamiltonian ($\widehat{H}_{CF}$) and the exchange coupling term resulted from the magnetization of Co atoms ($\widehat{H}_{ex}$):

$$\widehat{H}_{at} = \widehat{H}_U + \lambda \sum_i \hat{s}_i \hat{l}_i + \widehat{H}_{CF} + \widehat{H}_{ex}.$$

The eigenvalue of $\widehat{H}_{at}$ ($E_{eigen}$) was solved using the Lanczos algorithm as implemented in the Quanty code,[66] which corresponds to the magnetic anisotropy energy (defined as $E_{eigen} - E_{eigen}^{min}$) by varying the magnetization direction represented by the polar angle $\theta$ ($[0, \pi]$) and the azimuth angle ($[-\pi, \pi]$). The $\widehat{H}_U$ was specified by the Slater parameters $F^2 = 11.93, F^4 = 7.49$ and $F^6 = 5.39$, which were taken from Ref. [67]. $F^0$ follows the relation that $F^0 = U + F^2 * 4/195 + F^4 * 2/143 + F^6 * 100/5577$. Here, we simply took $U = 0$ since it is the relative energy change with respect to magnetization direction of more interest rather than the absolute eigenvalue of the Hamiltonian. Similarly, for $\widehat{H}_{CF}$ the $B_0^0$ term was also omitted. The SOC strength $\lambda$ of $Tb^{3+}$ atom was set to 0.221.[67] For $\widehat{H}_{ex} = 2\mu_B B_{ex} \boldsymbol{n} \cdot \widehat{\boldsymbol{S}}_f$ ($\widehat{\boldsymbol{S}}_f$ is the spin of Tb 4*f* shell), we assumed the magnitude of exchange field $B_{ex}$ to be 310 T as reported in the crystalline $TbCo_5$.[68] By solving the eigenfunctions of the Hamiltonian, we also obtained the expectation value of the magnetic dipole term defined as $\widehat{\boldsymbol{T}} = \widehat{\boldsymbol{S}} - 3\widehat{\boldsymbol{r}}(\widehat{\boldsymbol{r}} \cdot \widehat{\boldsymbol{S}})$, in which $\widehat{\boldsymbol{S}}$ and $\widehat{\boldsymbol{r}}$ are spin and radial position operators, respectively. In addition, the variation of the eigenvalue as a function of $(\theta, \varphi)$ was examined under different $B_{ex}$, as illustrated in Fig. S12. Clearly, despite the distinct energy levels, the shape of the curve, i.e., the energetically favoured magnetization direction, is not influenced by the magnitude of $B_{ex}$. We also adopted the CFPs with $\Delta = -0.4$ Ry for Co 3$d$ states in Tb25Co125 for the evaluation of magnetic anisotropy, i.e., assumed stronger hybridization between Tb and Co. Similarly, the distributions of high/low energy zones are not affected.

For each Tb atom in the pristine and hydrogen-charged amorphous structures, we generate a contour plot for $E_{aniso}$ as a function of $\theta$ and $\varphi$ (see the videos in SI for Tb25Co125, Tb25H25Co125, Tb32Co118 and Tb32H32Co118.). Centred at each Tb atom, its local chemical environment is represented by Tb-Co and Tb-Tb, and/or Tb-H bonding within cut-off radii of 3.5, 4.0 and 3.0 Å, respectively. Figure S12 offers a better visualization of the distributions of $E_{aniso}$ in Tb25Co125 and Tb32Co118 along the directions corresponding to various chemical bonding. since the distributions of $E_{aniso}$ corresponding to Tb-Co bonding directions below the middle bar $(E_{aniso}^{min} + E_{aniso}^{max})/2$ are much denser than that of above the middle bar. In addition, for a larger fraction of Tb atoms, the Tb-Tb bonding directions either locate right along or close to the hard magnetization directions.

The CFPs of Tb12 in both pristine and hydrogenated structures are listed in Table S2. In order to investigate which parameter is mainly responsible for the change of anisotropy displayed in Fig. 4 D and E, we have separately checked how the variations of dominating CFPs $B_2^0, B_4^0, B_6^0$, and $B_6^6$ influence the shape of the magnetic anisotropy. In specific, as for Tb12 in Tb25Co125, the $B_2^0, B_4^0, B_6^0$, and $B_6^6$ are individually modified to be the same as those in $TbHCo_5$ while the other CFPs are fixed. As demonstrated in Figure S13, by comparison, the change of the $E_{eigen}$ curve caused by hydrogen addition is mostly represented by the modification of $B_2^0$ term, whereas the variations of $B_4^0, B_6^0$, and $B_6^6$ exhibit negligible contributions. Figure S14 demonstrates a more direct comparison between $E_{aniso}$ with negative and positive $B_2^0$ while leaving out all other higher-order terms. Here, we force the same magnitude of $B_2^0$ while change the sign since the magnitude affects only the size of $E_{aniso}$.It shows that a 90° reorientation of the easy magnetization axis from ***c*** to ***a*** axis in the simulated structure.

**Evaluation of strain effect on anisotropy switching**

Magnetoelastic effect may also be involved at the Pd/TbCo and TbCo/substrate interface due to hydrogen insertion. Hydrogen absorption in Pd layer induces the lattice expansion up to 4%[69] and subsequently TbCo layer can experience elastic strain. However, in our case we want to emphasize that the thickness of Pd layer is substantially smaller than that of TbCo layers. We have observed the hydrogen-induced anisotropy switching in Pd (2nm)/TbCo (20 nm, 50 nm) layers, and the thickness of Pd layer is only 2 nm in comparison with 50 nm thickness of TbCo layer. Hence, it is unlikely that the lattice expansion of Pd will affect the bulk anisotropy of TbCo layer. Another direct evidence is that, when Pd layer was discharged by applying 0.2 V (desorption of hydrogen), magnetic anisotropy of TbCo layer remained in-plane. This observation excludes the effect of lattice expansion in Pd layer as the origin of anisotropy switching in amorphous TbCo.

We next consider the elastic strain between TbCo/substrate, elastic strain can exist between TbCo layer and substrate. The lattice volume of TbCo layer may change much less upon hydrogen insertion due to large interstitial space in amorphous structure.[70] Furthermore, the transfer of elastic strain between substrate and thin films usually require high-quality interface (epitaxial growth of thin films) and small thickness of thin films (for instance below 10 nm). However, the conditions of our sputtering process, which was conducted at room temperature, resulted in a rough interface for without epitaxial growth. The thickness of the TbCo layer, reaching 50 nm, further decreases the likelihood of significant strain transfer. In addition, to better address the strain effect, we performed ab initio calculations to investigate the change of magnetic anisotropy using an elongated atomic structure along $c$ direction, which simulates the condition that the in-plane lattice is constraint that leaves only the out-of-plane degree of freedom. The strain level was set to 2% and 5% in the ab initio calculations, respectively. The results show that the magnitudes of crystal-field parameters (CFPs) are altered by around 0.001 eV (less than 5% of the CFPs before structural elongation) under strain up to 5%. As exemplified in Fig. S16, the contour map of $\mathrm{E_{aniso}}$ demonstrate negligible changes due to the strain effect. All these considerations and results lead us to believe that the strain effects may not play a substantial role in inducing anisotropy change.

**ASSOCIATED CONTENT**

The Supporting Information is available free of charge online.

Hysteresis loops of anomalous Hall resistance at various temperatures for TbCo films with original $T_M$ of 198 K, 295 K before and after applying -1.2 V (Fig. S1, S2, S3); Magnetic hysteresis loops for TbCo films with $T_M$ of 198 K at various temperatures before and after applying -1.2 V (Fig. S4); Microstructure characterization and element distribution of TbCo thin films (Fig. S5-8); Contour diagram of the expectation value for the magnetic dipole operator $\langle T_z \rangle$ as a function of the polar angle θ ([0, π]) and the azimuth angle $\phi$ ([-π,π]) (Fig. S9); Dependence of the orbital, spin, and total magnetic moments on the expectation values $\langle \boldsymbol{T_z} \rangle$ values (Fig. S10); Step functions for Co (left, red curve) and for Tb (right, red curve) used to account for the transitions into the continuum (Fig. S11); The TbCo supercell built to simulate amorphous structure (Fig. S12); The distributions of $E_{aniso}$ at $(\theta, \varphi)$ corresponding to the Tb-Co (marked by red cross) and Tb-Tb (marked by purple diamonds) bonds within the cluster for each Tb atom (Fig. S13); $E_{eigen}$ of Tb12 as a function of magnetization direction ($E_{eigen}(0, -\pi)$ as the reference (Fig. S14) and its contour plot (Fig. S15); Contour maps of $\mathrm{E_{aniso}}$ plotted against θ and φ for Tb19 (a) without and (b) with strain (Fig. S16). The resolved spin and orbital magnetic moments of Co and Tb in Tb-o films in units of $\mu_B$/atom along grazing and perpendicular directions before and after hydrogen insertion (Table S1); Crystal field parameters (CFPs) of Tb12 in amorphous Tb25Co125 and Tb25H25Co125 (Table S2).

**Figures and figure captions**

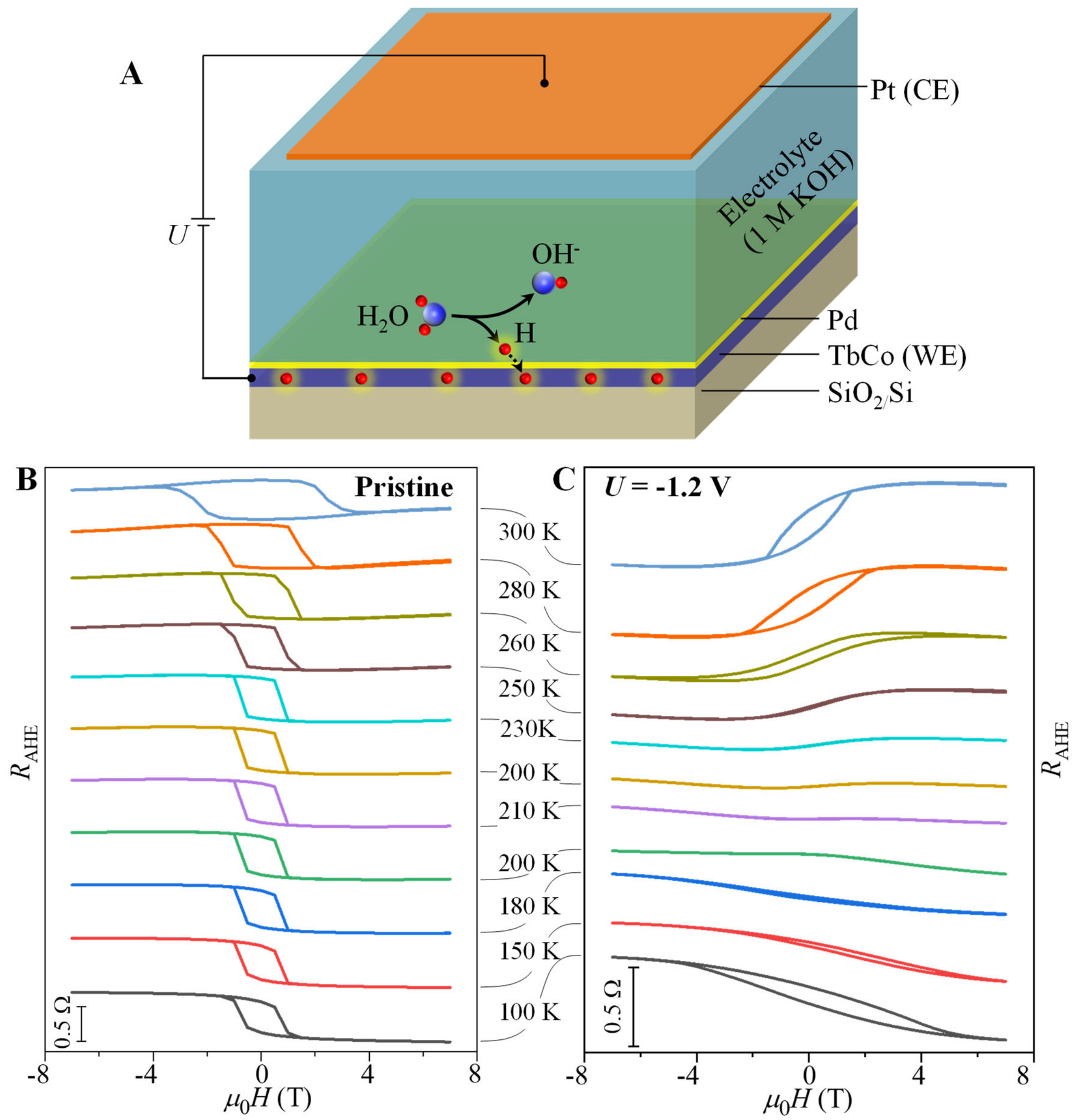

**Fig. 1 Voltage-driven 90° switching of bulk PMA to in-plane directions in Tb-Co ferrimagnetic films ($T_M$ > 300 K).** (A) Schematic of an electrochemical setup used for voltage-driven hydrogen insertion and the sample configuration of Pd/TbCo/$SiO_2$/Si. When -1.2 V is applied between the working electrode (WE, TbCo layer) and the counter electrode (CE, Pt foil) across the electrolyte (1 M KOH aqueous solution), water molecules will be reduced into $OH^-$ and hydrogen atoms (H) on the surface of Pd layers. Driven by gradient of concentration, hydrogen atoms will sequentially diffuse into Pd and TbCo layers. (B) Hysteresis loops of anomalous Hall resistance for as-grown thin films and (C) for those after applying -1.2 V, measured at temperatures from 300 K to 100 K. Note the 90° switching of magnetic anisotropy and the reduction of $T_M$ from initially above 300 K to about 210 K after hydrogen insertion. Magnetic fields were applied along out-of-plane directions.

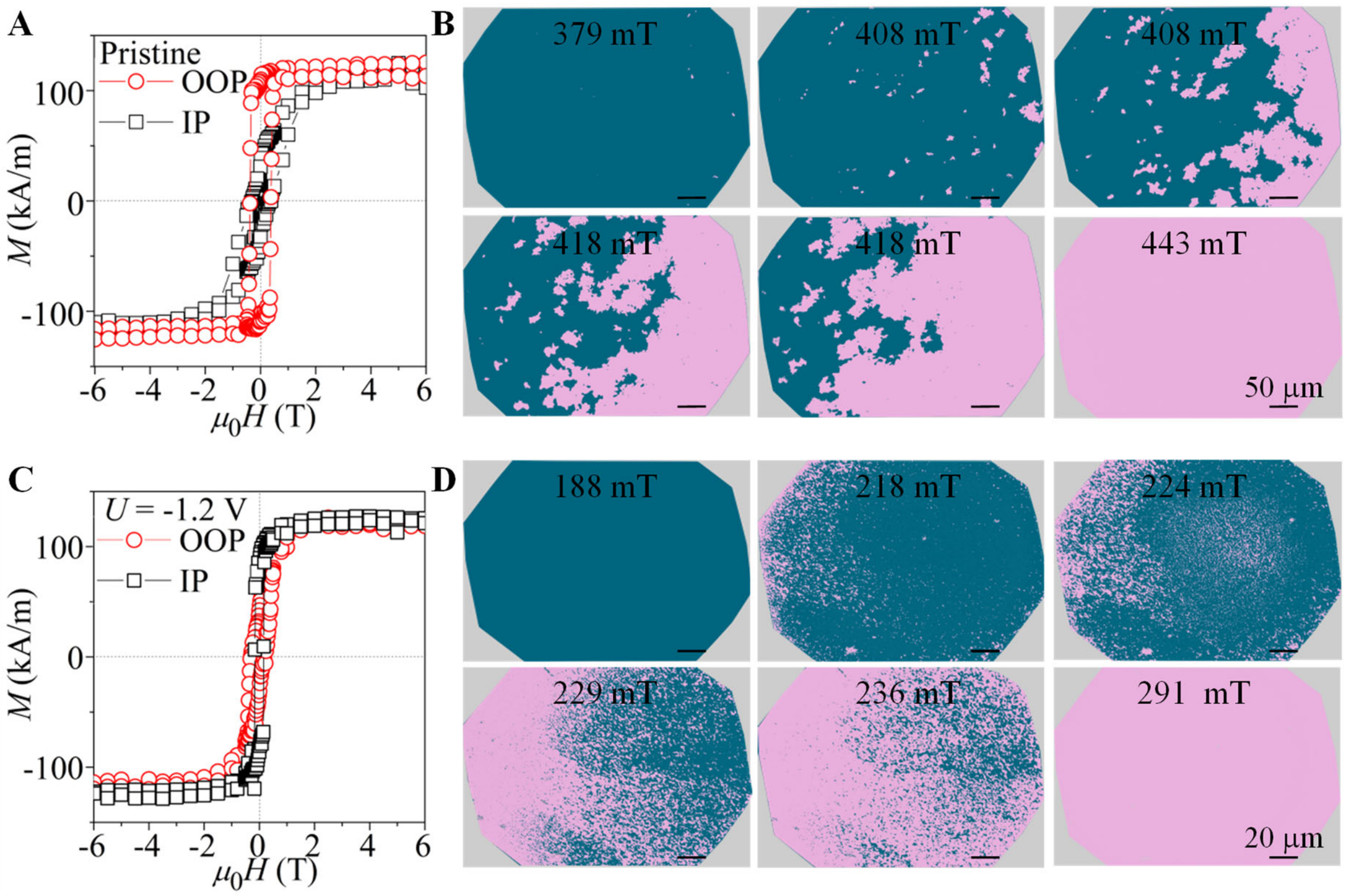

**Fig. 2 Magnetization behavior in Tb-Co ferrimagnetic films ($T_M$ ~ 200 K) before and after applying -1.2 V.** (A) Magnetic hysteresis loops of as-grown films measured at room temperature along out-of-plane and in-plane directions, evidencing the robust PMA. (B) The corresponding MOKE images of magnetic domain structure under increasing magnetic fields applied along out-of-plane directions. Note the nucleation and displacement of magnetic domain walls under constant magnetic fields of 408 and 418 mT. (C) Magnetic hysteresis loops of TbCo films after the application of -1.2 V at room temperature, demonstrating 90° anisotropy switching in comparison with (A). (D) The corresponding MOKE images of magnetic domain structure after applying -1.2 V, showing the gradual formation of isolated, interspersed domains. Low-temperature magnetic hysteresis loops are shown in Fig. S4.

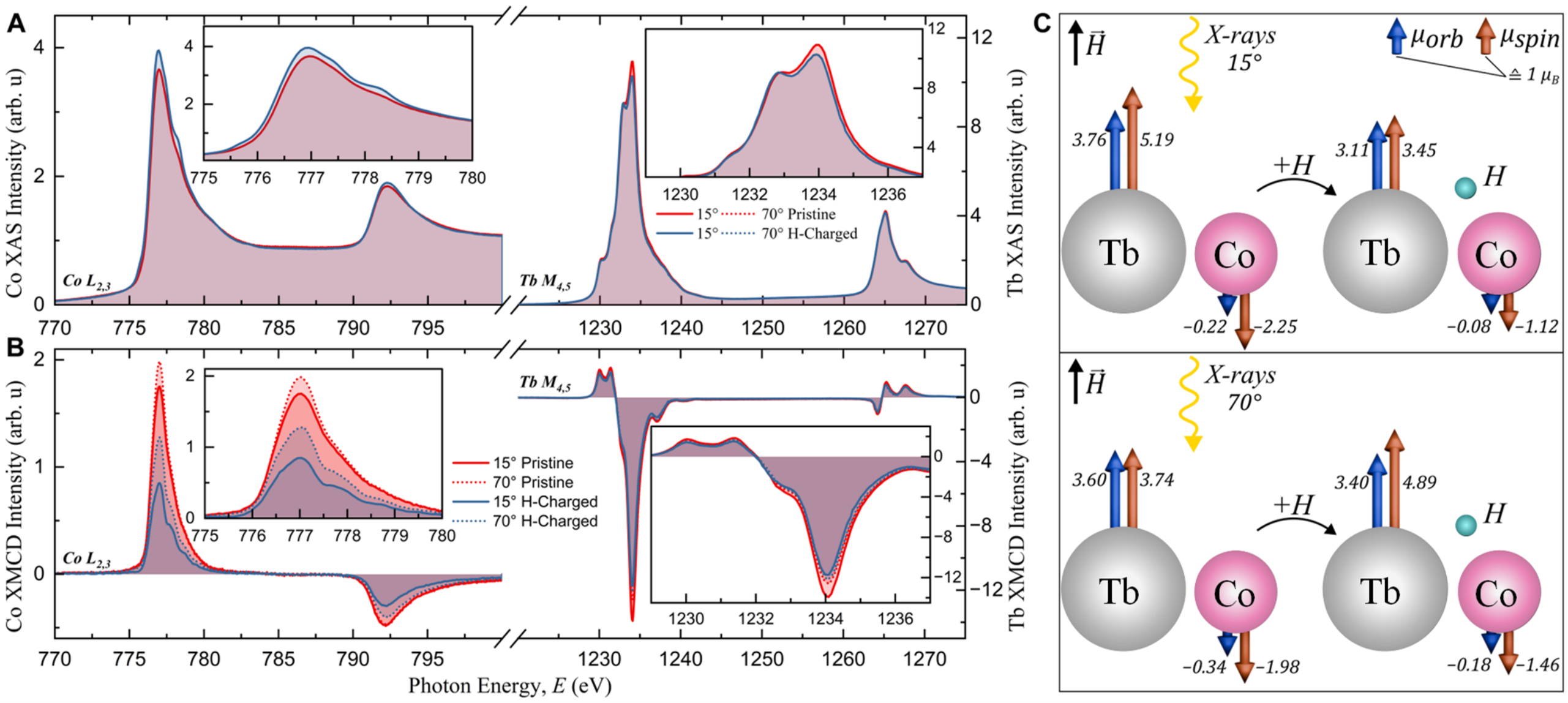

**Fig. 3 Angle-dependent XAS and XMCD spectra for TbCo ferrimagnetic films ($T_M$~295 K) before (pristine) and after applying -1.2 V (H-Charged).** (A) XAS spectra at Tb $M_{4,5}$ and Co $L_{2,3}$ edges for samples before and after hydrogen insertion, which resemble each other in terms of peak positions and line shapes. (B) The corresponding XMCD spectra for Co and Tb along both perpendicular and grazing incidence angles, showing the substantial dependence on incidence angles in accordance with magnetic anisotropy. (C) The resolved spin and orbital moments of Co and Tb by sum rules, showing the opposite dependence of Tb orbital moments on incidence angles. The given numbers of the moments are in the units of $\mu_B$/atom. All measurements were performed at $T$ = 5 K, with an applied magnetic field of 8 T parallel to the incoming photon beam, and perpendicular and grazing angles are 15° and 70° with respect to the surface normal of thin films.

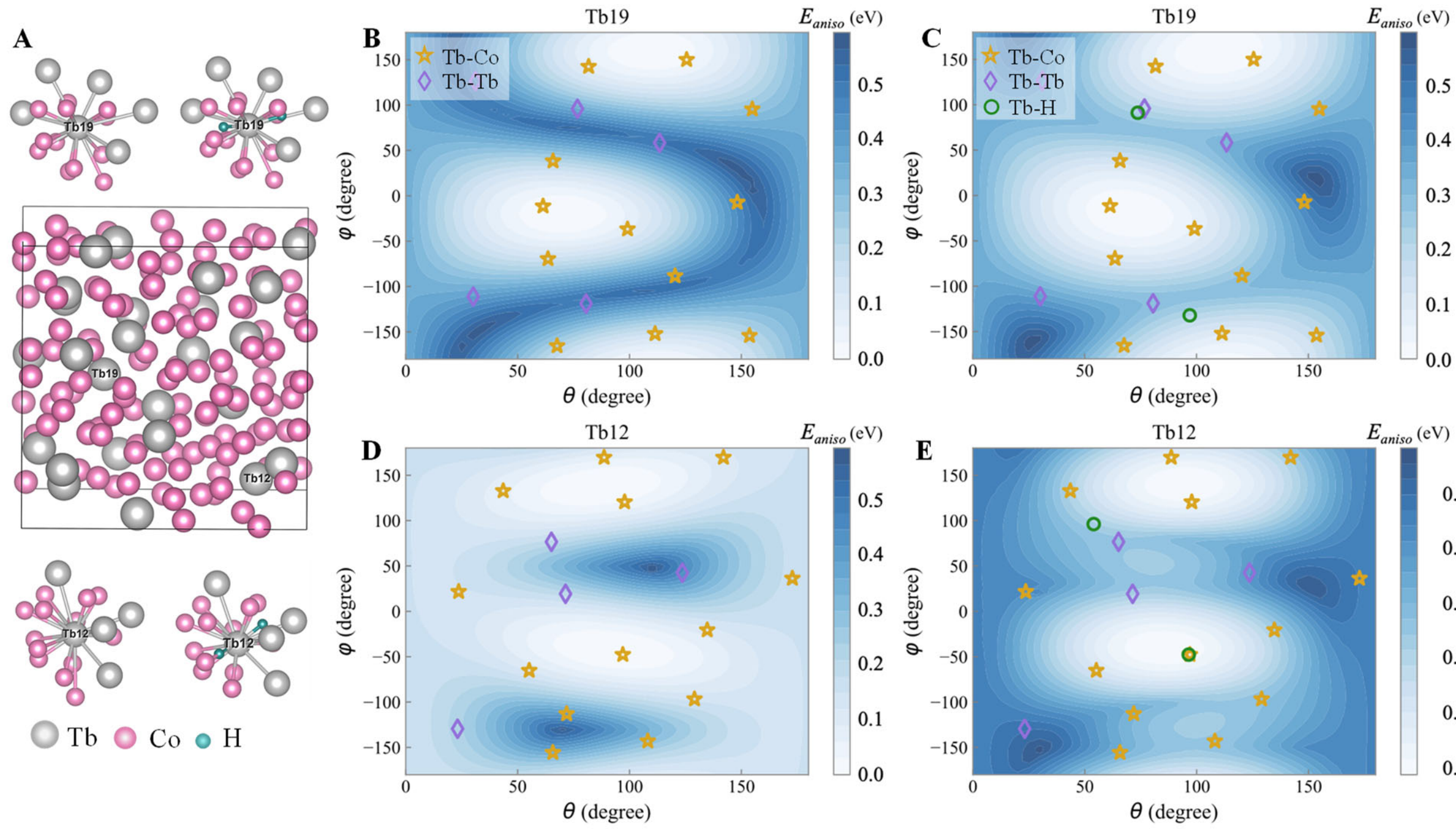

**Fig. 4 Ab initio calculations of three-dimensional magnetic anisotropy contour in TbCo atom configuration before and after hydrogen insertion.** (A) The supercell structure to simulate the amorphous structure of TbCo built through stochastic quenching. The spatial directions and vectors of Tb-Tb and Tb-Co bonding were projected in spherical coordinate with polar and azimuth angles $(\theta, \varphi)$. Top and bottom panels show two representative clusters centered around Tb atoms (Tb12 and Tb19). (B) Contour maps of $E_{\text{aniso}}$ plotted against $\theta$ and $\varphi$ for Tb19, showing the high anisotropy energy along Tb-Tb direction (diamond symbols on the blue arc). (C) Contour maps of $E_{\text{aniso}}$ plotted against $\theta$ and $\varphi$ for Tb19 after hydrogen insertion, showing the substantial weakening of the anisotropy difference between Tb-Co and Tb-Tb bonding directions. (D) Contour plot of magnetic anisotropy energy around Tb12 before and (E) after hydrogen insertion, evidencing the 90° reorientation of anisotropy direction due to hydrogen insertion.

## Contributions

XY conceptualized and supervised the work. ZX prepared thin films, conducted AHE and magnetometry measurements and performed electrochemical measurements under the supervision of XY. RX, HZ carried out the DFT calculations. FM, XY performed MOKE characterization. PK, BE, TH, JL, and KK performed the XMCD measurements at the ESRF, and PK, BE analysed the XMCD data under the supervision of HW and KO. YD prepared TEM samples and DW performed TEM characterizations. XL wrote the manuscript and coordinated the collaborations among the groups in KIT, TU Darmstadt and University of Duisburg-Essen. All authors contributed to the discussion of the results and the revision of the manuscript.

## Acknowledgement

XL acknowledge financial support by the Deutsche Forschungsgemeinschaft (DFG) under grant number 528530757. OG, KS, HZ, HW, acknowledge financial support by the Deutsche Forschungsgemeinschaft (DFG) within the CRC/TRR 270 (Project-ID 405553726). PK, BE acknowledge the financial support from German Federal Ministry of Education and Research (BMBF) under Grant BMBF-Projekt05K2022. We acknowledge the ESRF for the allocation of beam time at ID32 under proposal number IH-HC-3833 and the Karlsruhe Nano Micro Facility (KNMF). We also acknowledge the computing time provided at the NHR Center NHR4CES at RWTH Aachen University (project number p0020449) and at TU Darmstadt (project number p0020538). It is funded by the Federal Ministry of Education and Research, and the state governments participating on the basis of the resolutions of the GWK for national high performance computing at universities.

Supplementary materials for

# Voltage-gated 90° switching of bulk perpendicular magnetic anisotropy in ferrimagnets

Zhengyu Xiao[1,2,3♪], Ruiwen Xie[4♪], Fernando Maccari[4], Philipp Klaßen[5], Benedikt Eggert[5], Di Wang[2,6], Yuting Dai[2], Raquel Lizárraga[7,8], Johanna Lill[5], Tom Helbig[5], Heiko Wende[5], Kurt Kummer[9], Katharina Ollefs[5], Konstantin P Skokov[4], Hongbin Zhang[4], Zhiyong Quan[3,10], Xiaohong Xu[3,10], Robert Kruk[2], Horst Hahn[2,11], Oliver Gutfleisch[4],Xinglong Ye[4,1,12*]

[1]TUD-KIT Joint Research Laboratory Nanomaterials, Technische Universität Darmstadt, 64287 Darmstadt, Germany
[2]Institute of Nanotechnology, Karlsruhe Institute of Technology, 76344 Eggenstein-Leopoldshafen, Germany
[3]Key Laboratory of Magnetic Molecules and Magnetic Information Materials of Ministry of Education, School of Chemistry and Materials Science, Shanxi Normal University, 030032 Taiyuan, China
[4]Institute of Materials Science, Technische Universität Darmstadt, 64287 Darmstadt, Germany
[5]Faculty of Physics and Center for Nanointegration Duisburg-Essen (CENIDE), University of Duisburg-Essen,
47057 Duisburg, Germany
[6]Karlsruhe Nano Micro Facility, Karlsruhe Institute of Technology (KIT), 76131 Karlsruhe, Germany
[7]Applied Materials Physics, Department of Materials Science and Engineering, Royal Institute of Technology, Stockholm SE-100 44, Sweden
[8]Wallenberg Initiative Materials Science for Sustainability, Royal Institute of Technology, Stockholm SE-100 44, Sweden
[9]European Synchrotron Radiation Facility, BP 220, F-38043, Grenoble Cedex, France
[10]Collaborative Innovation Center for Shanxi Advanced Permanent Magnetic Materials and Technology, Research Institute of Materials Science, Shanxi Normal University, 030032 Taiyuan, China
[11]The University of Oklahoma, School of Sustainable Chemical, Biological and Materials Engineering, Norman OK 73019, United States
[12]Present address: School of Physics, Shandong University, 250100 Jinan, China

♪ These authors contributed equally to this work.
* xinglong.ye@tu-darmstadt.de

**This PDF file includes:**

**Fig. S1 to S16**

**Table S1, S2**

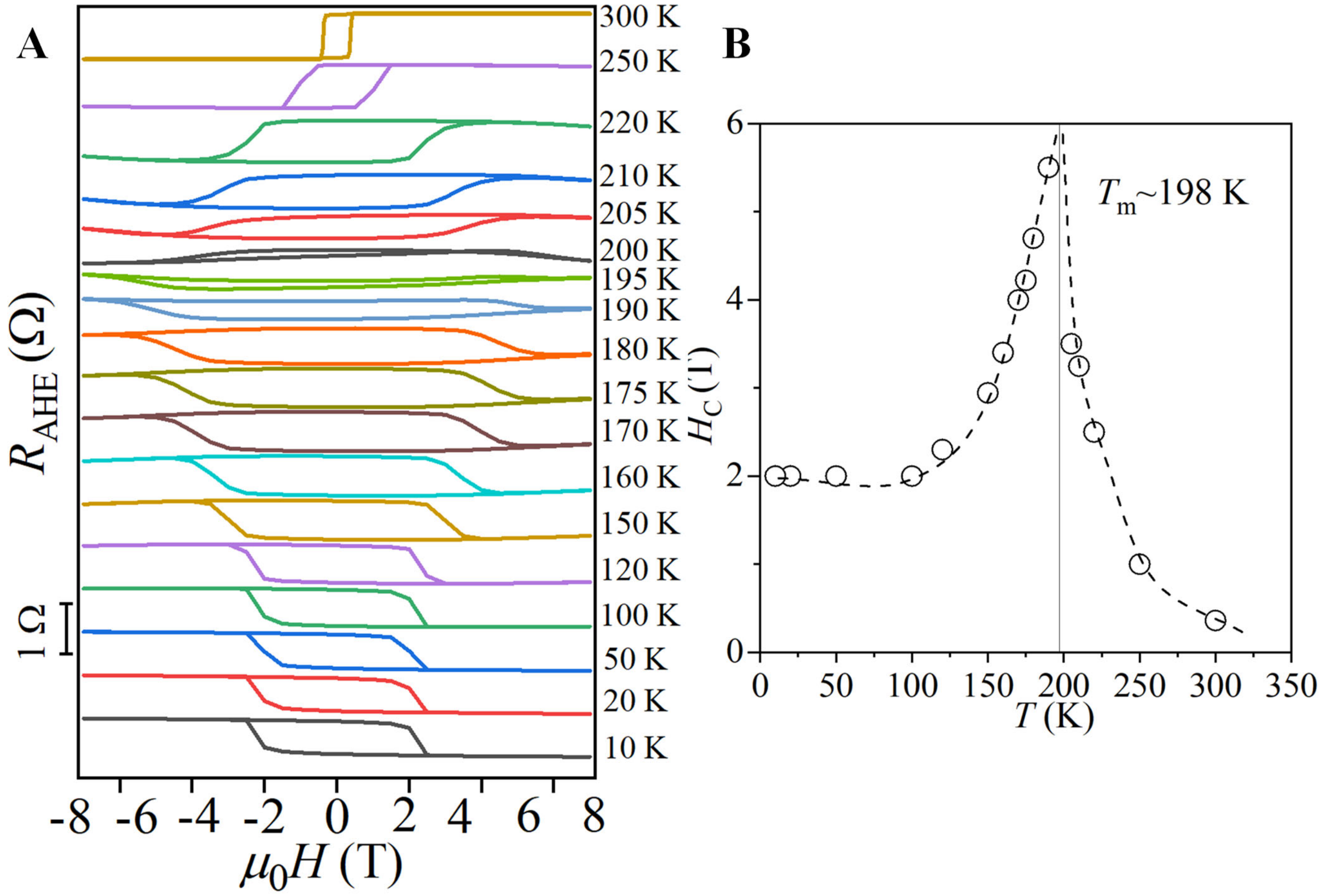

**Figure S1 Hysteresis loops probed by AHE for as-grown Tb-Co films in correspondence to the sample presented in Fig. 2.** (A) AHE loops measured in temperatures ranging from 10 K to 300 K, showing the polarity inversion at $T_M$ around 200 K. (B) Temperature dependence of coercivity obtained from (A), showing the divergence of coercivity at $T_M$ of 198 K.

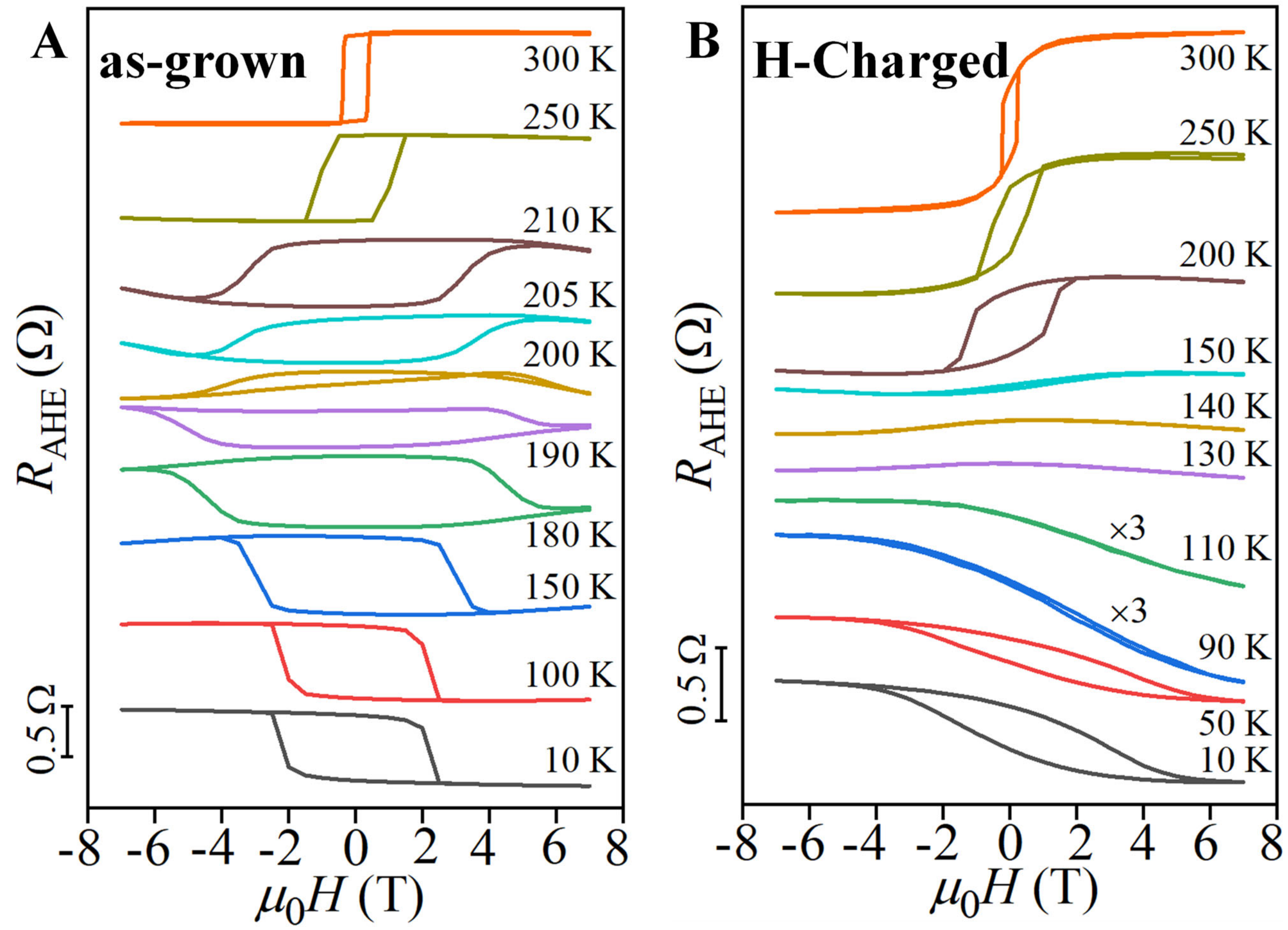

**Figure S2 Voltage-driven switching of magnetic anisotropy from perpendicular to in-plane directions in Tb-Co films with $T_M$ of 198 K.** (A) AHE loops taken from Fig. S1, exhibiting typical rectangular shapes with large PMA. (B) AHE loops after hydrogen insertion, exhibiting in-plane anisotropy. It is also evident that after hydrogen charging, the $T_M$ decreased from 200 K to 130 K, as identified by the inversion of AHE polarity.

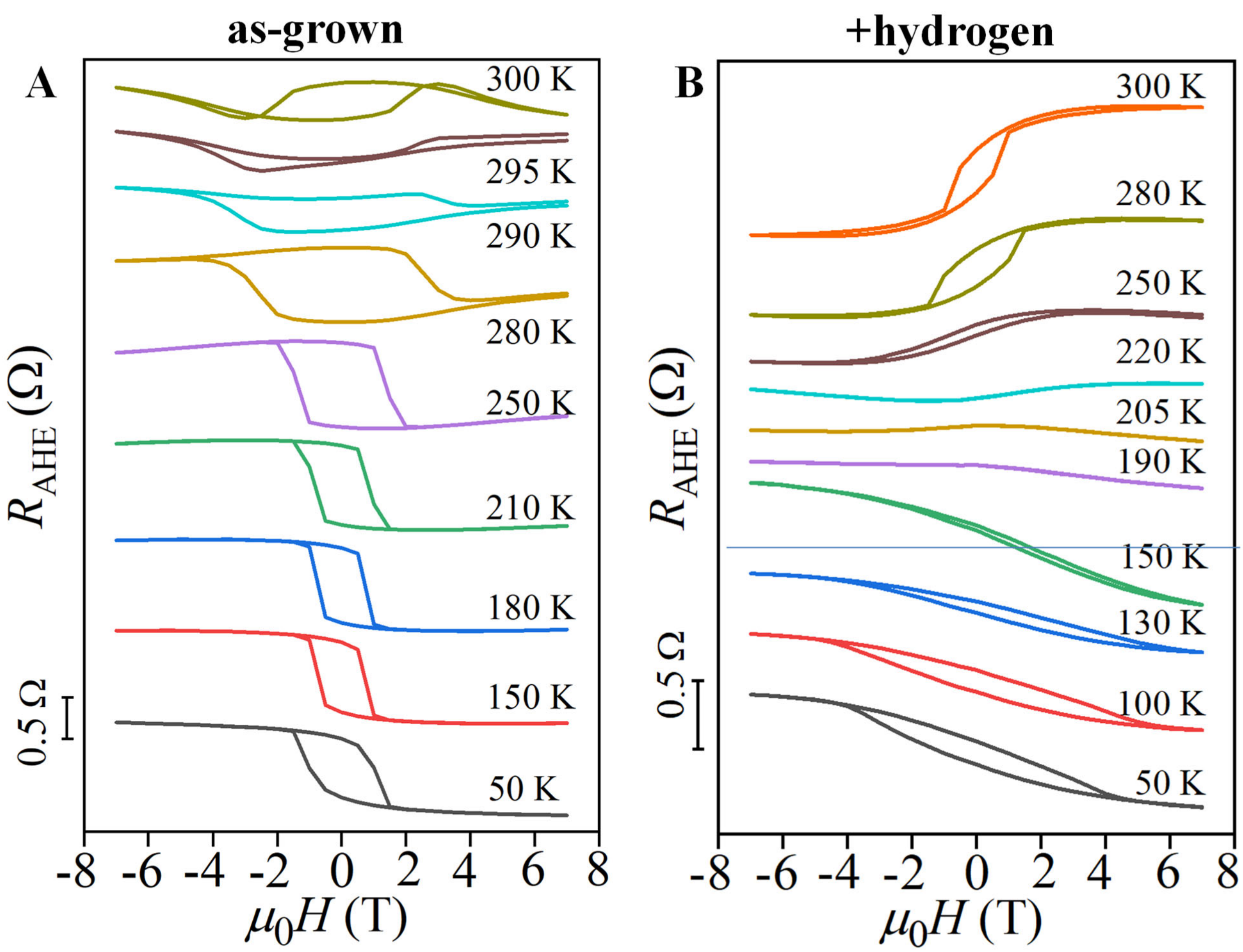

**Figure S3 Voltage-driven switching of magnetic anisotropy from perpendicular to in-plane directions in TbCo films with $T_M$ of 295 K in correspondence to the sample presented in Fig. 3.** (A) AHE loops of as-grown Tb-Co films with $T_M$ of 295 K, which shows typical rectangular shapes. (B) The corresponding AHE loops of TbCo thin films after hydrogen insertion, displaying a hard-axis loop with in-plane anisotropy. The $T_M$ decreased to approximately 205 K, as identified by the inversion of AHE polarity.

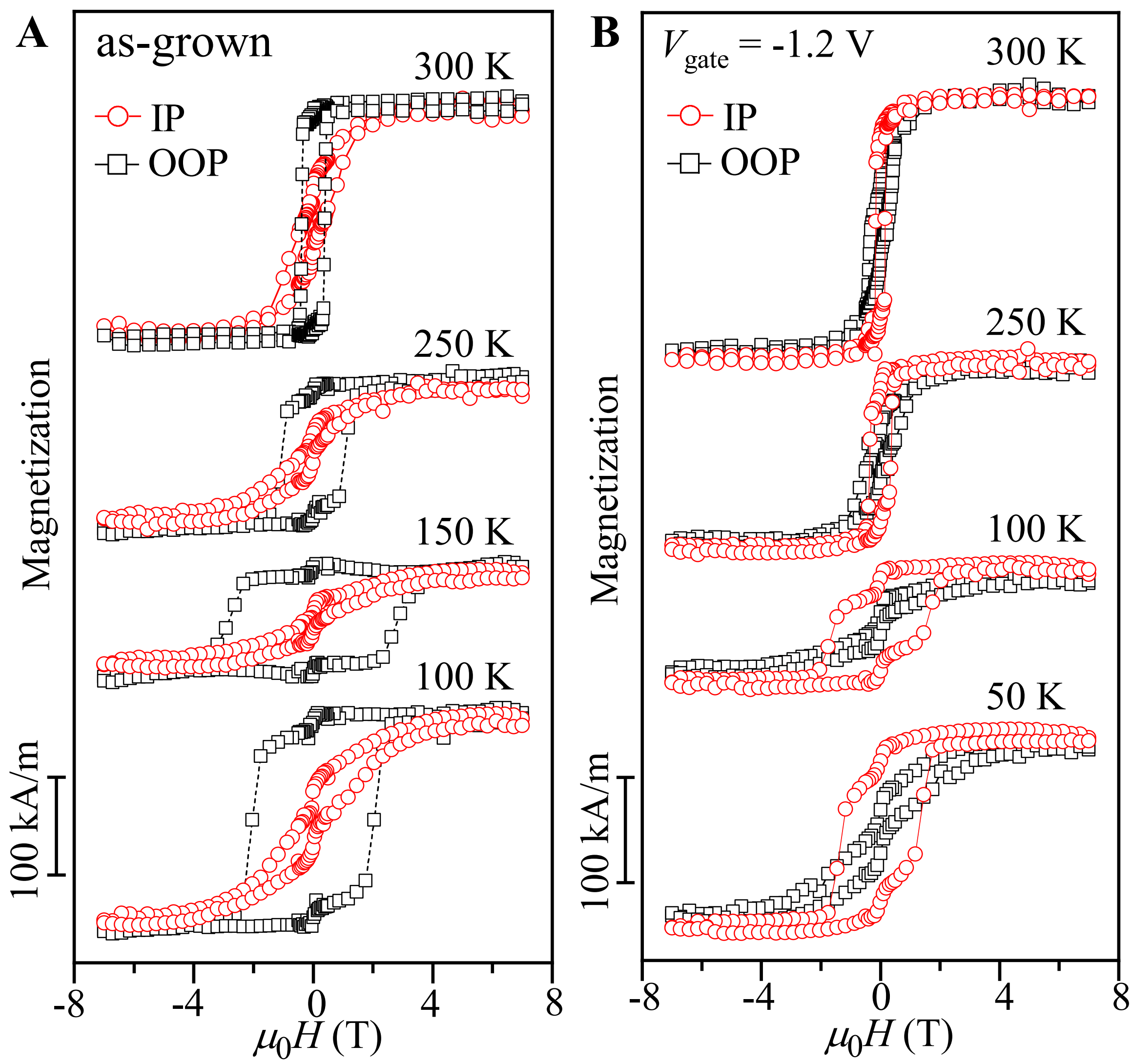

**Figure S4** Magnetic hysteresis loops of (A) as-grown films and those after applying -1.2 V measured at various temperatures along out-of-plane and in-plane directions, verifying the 90° switching from PMA to in-plane directions.

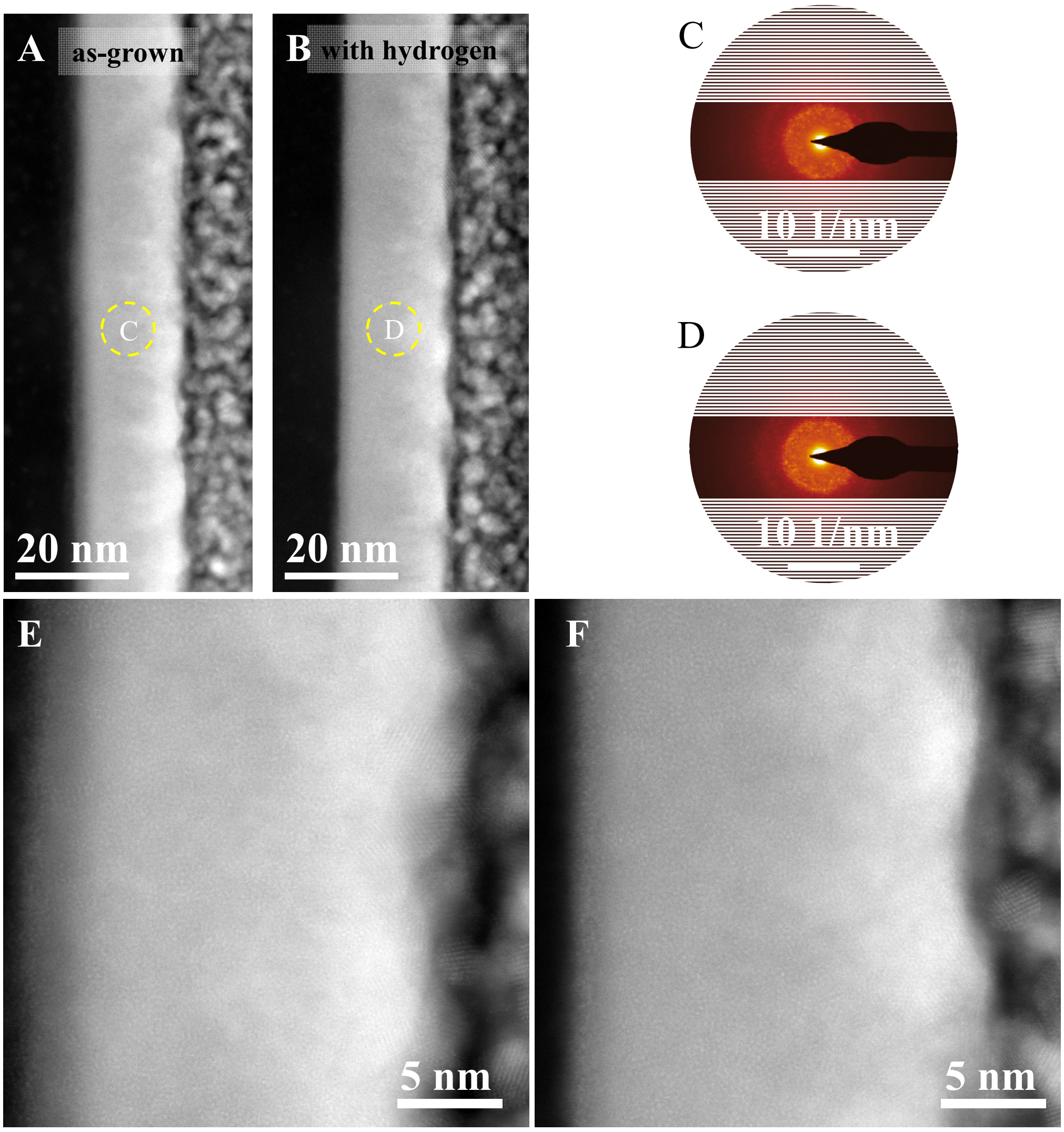

**Figure S5 TEM observation of the microstructure of TbCo thin films before and after voltage-driven hydrogen insertion.** STEM images of the cross section for (A) as-grown thin films and (B) those after hydrogen insertion. (C) The corresponding selected area diffraction (SAED) collected at the region in (A). (D) The corresponding SAED collected at region in (B). (E) A close-up HRTEM image of (A), showing amorphous structure without lattice fringes. (F) A close-up HRTEM image of B. Comparing the morphology and diffraction patterns reveal no discernable differences before and after hydrogen insertion.

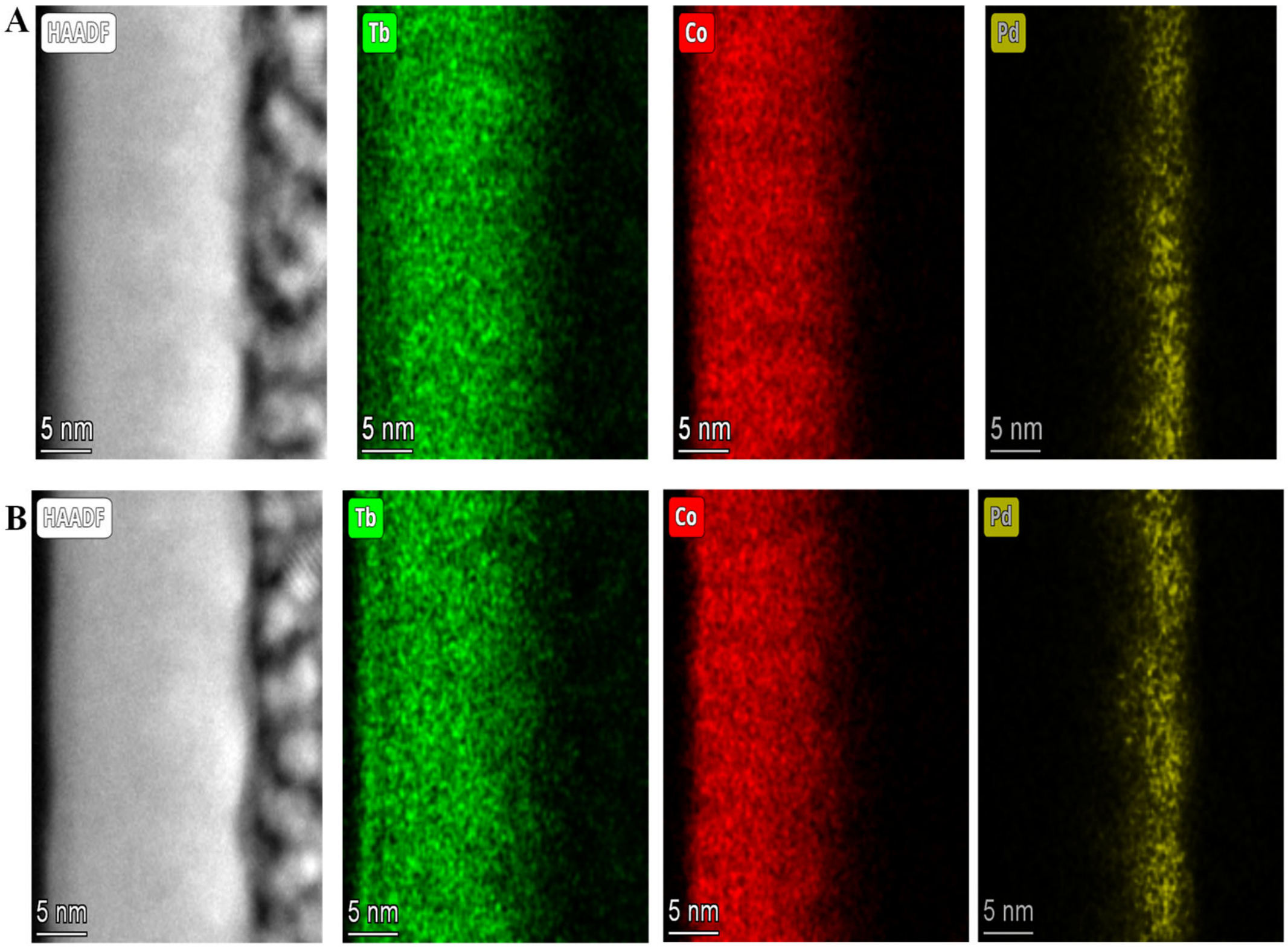

**Figure S6** EDS mapping of Tb, Co and Pd elements distribution over the cross section of TbCo thin films before (A) and after (B) voltage-driven hydrogen insertion, without detecting discernable differences.

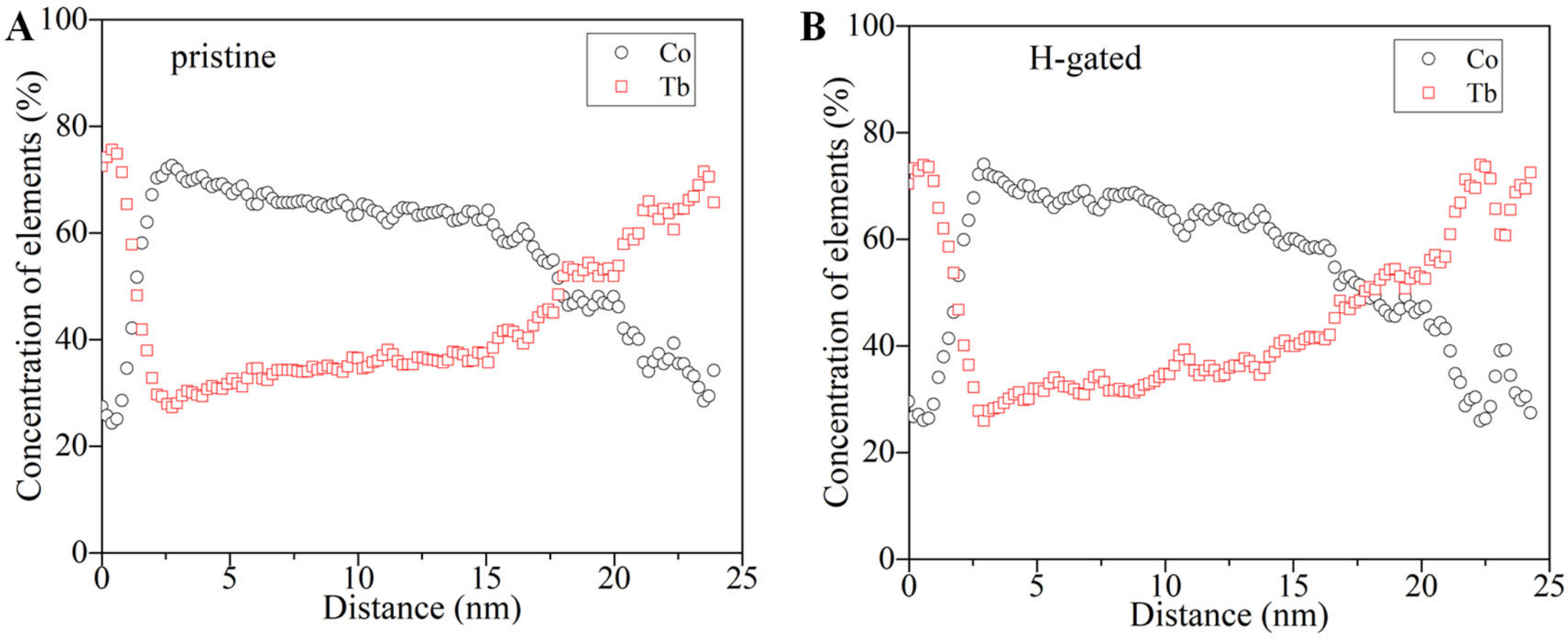

**Figure S7** Line profiles of Tb and Co distribution along perpendicular direction in TbCo thin films before (A) and after (B) hydrogen insertion, showing no discernable differences.

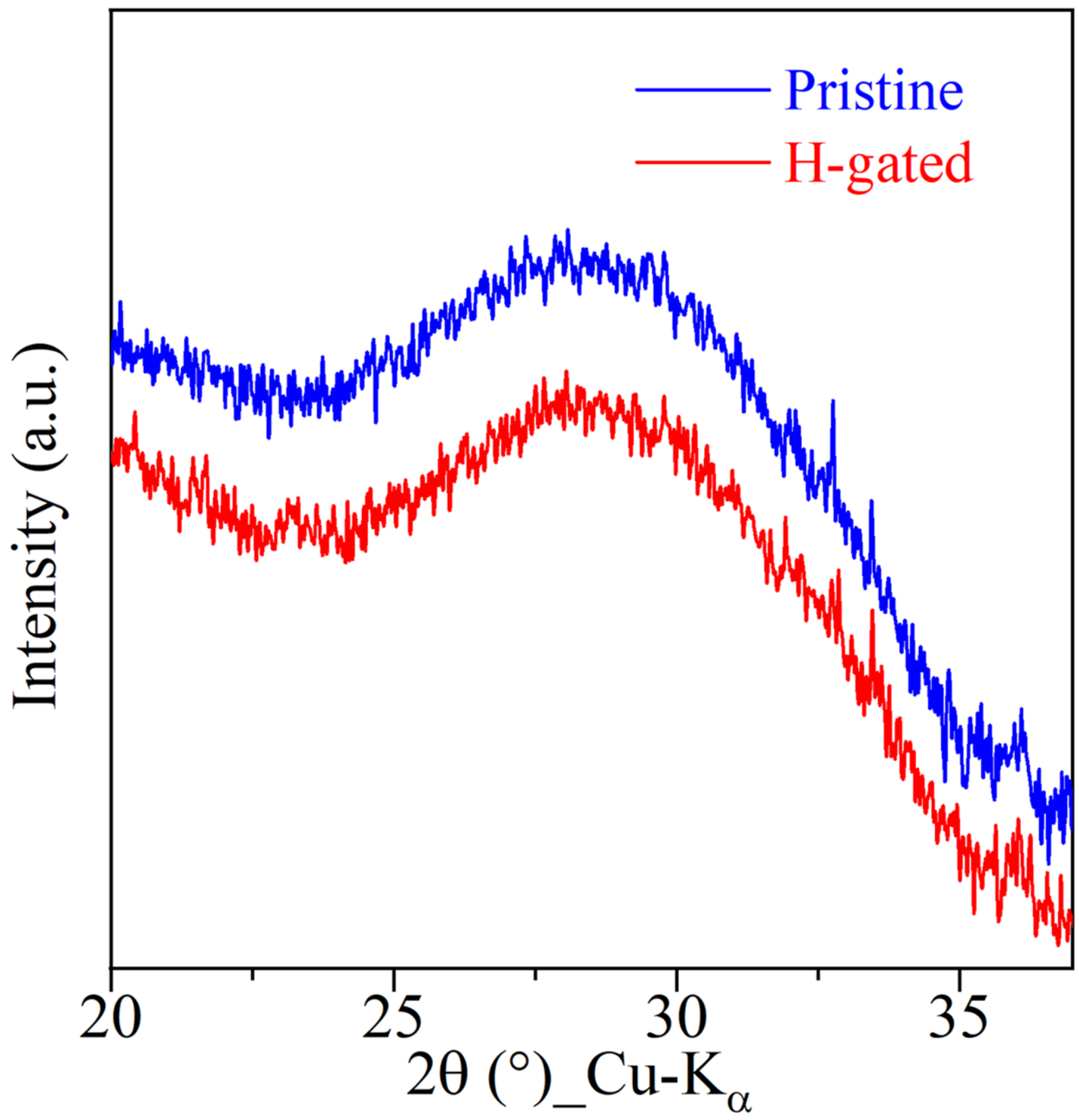

**Figure S8** XRD patterns of TbCo thin films before and after voltage-driven hydrogen insertion, confirming the maintenance of amorphous structure.

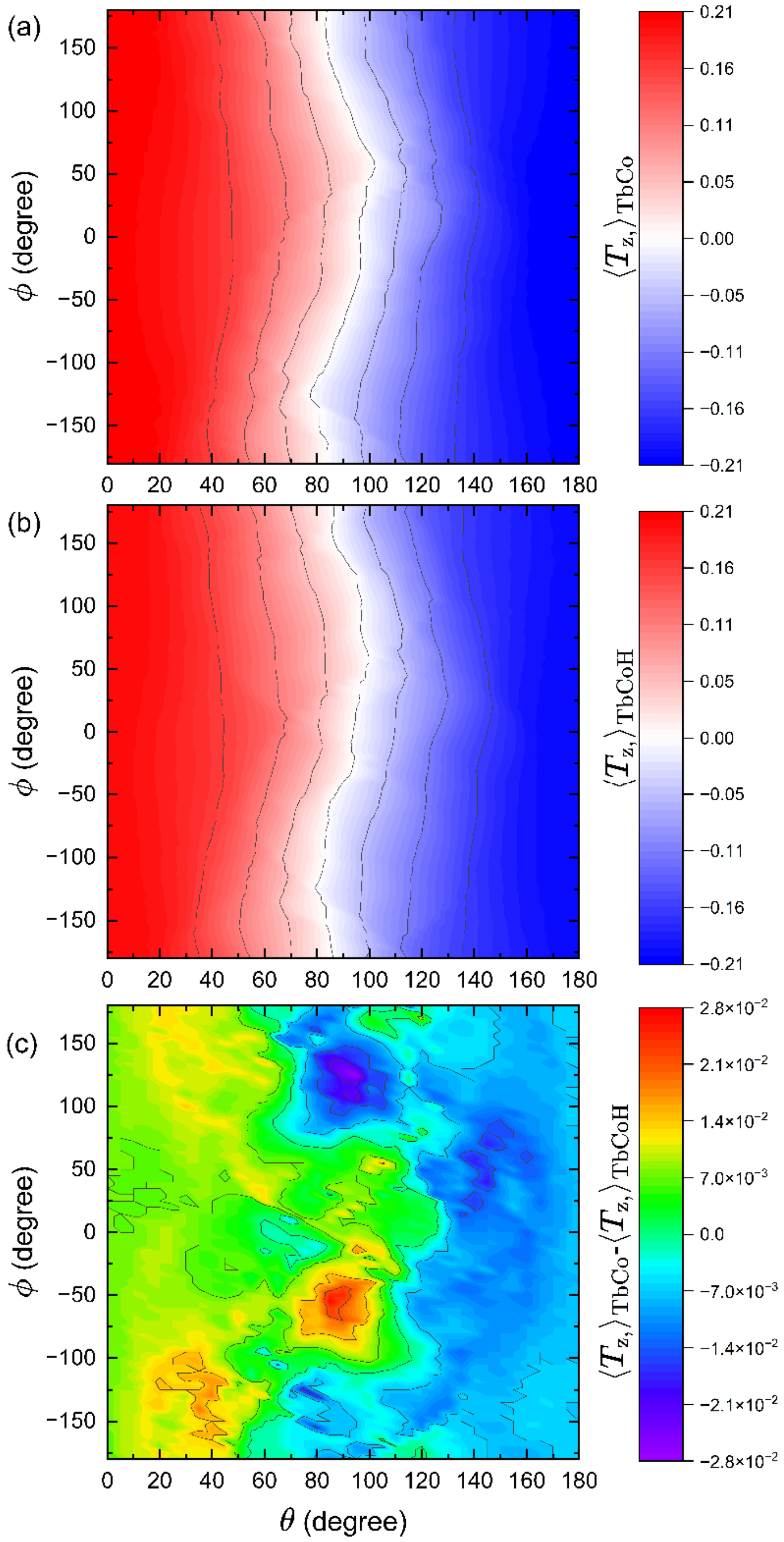

**Figure S9 Contour diagram of the expectation value for the magnetic dipole operator $\langle T_z \rangle$ as a function of the polar angle θ ([0, π]) and the azimuth angle $\phi$ ([-π,π]).** (A) The $\langle T_z \rangle$ values for as-grown samples and (B) for those after voltage-driven hydrogen insertion. (C) The difference between the two contour plots of $\langle T_z \rangle$ values.

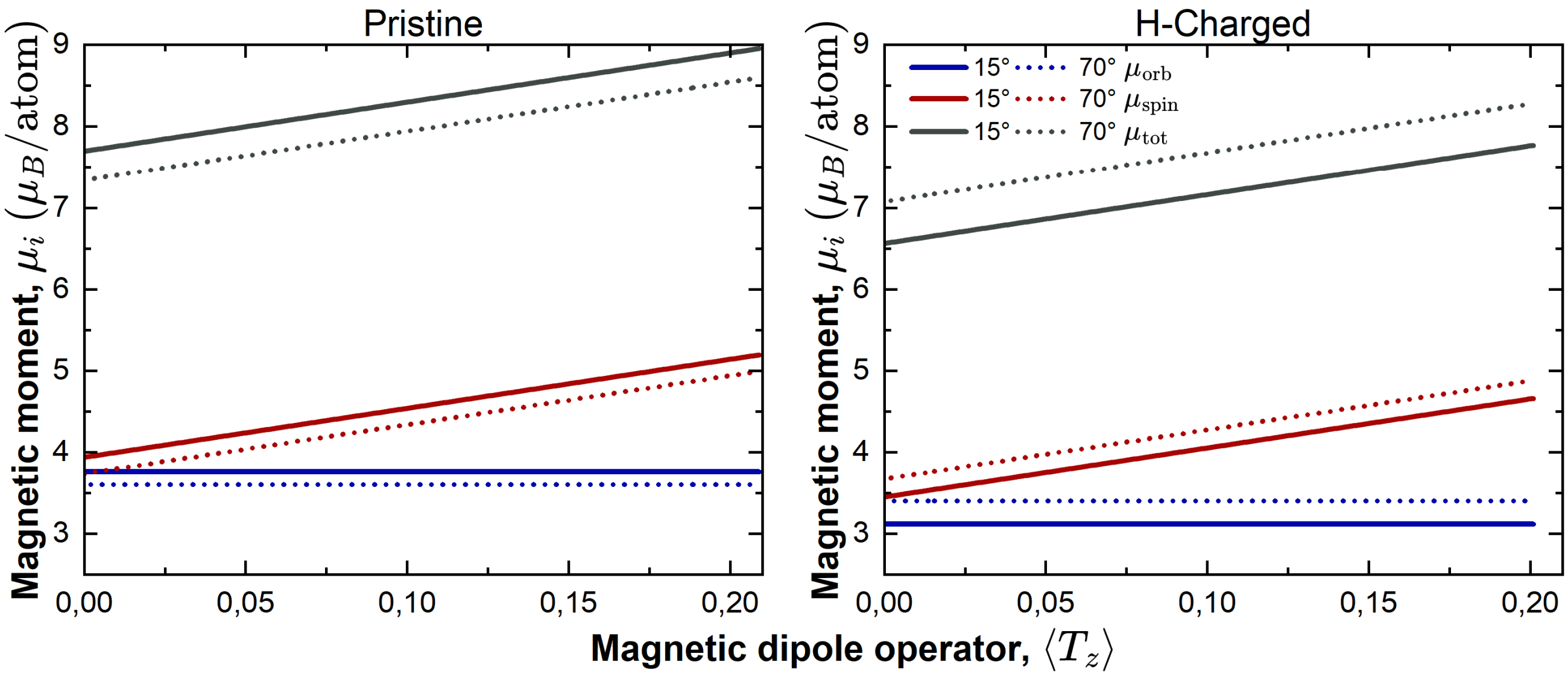

**Figure S10** Dependence of the orbital, spin, and total magnetic moments on the expectation values $\langle \boldsymbol{T_z} \rangle$ values for the range of the calculated terms using Quanty [1]. $\langle S_z \rangle$ and $\langle T_z \rangle$ are antiparallel for Tb3+ ions [2]. Given $\langle S_z \rangle$ negative, only positive values were considered. The results for the as-grown sample are shown on the left, while for that after hydrogen insertion on the right.

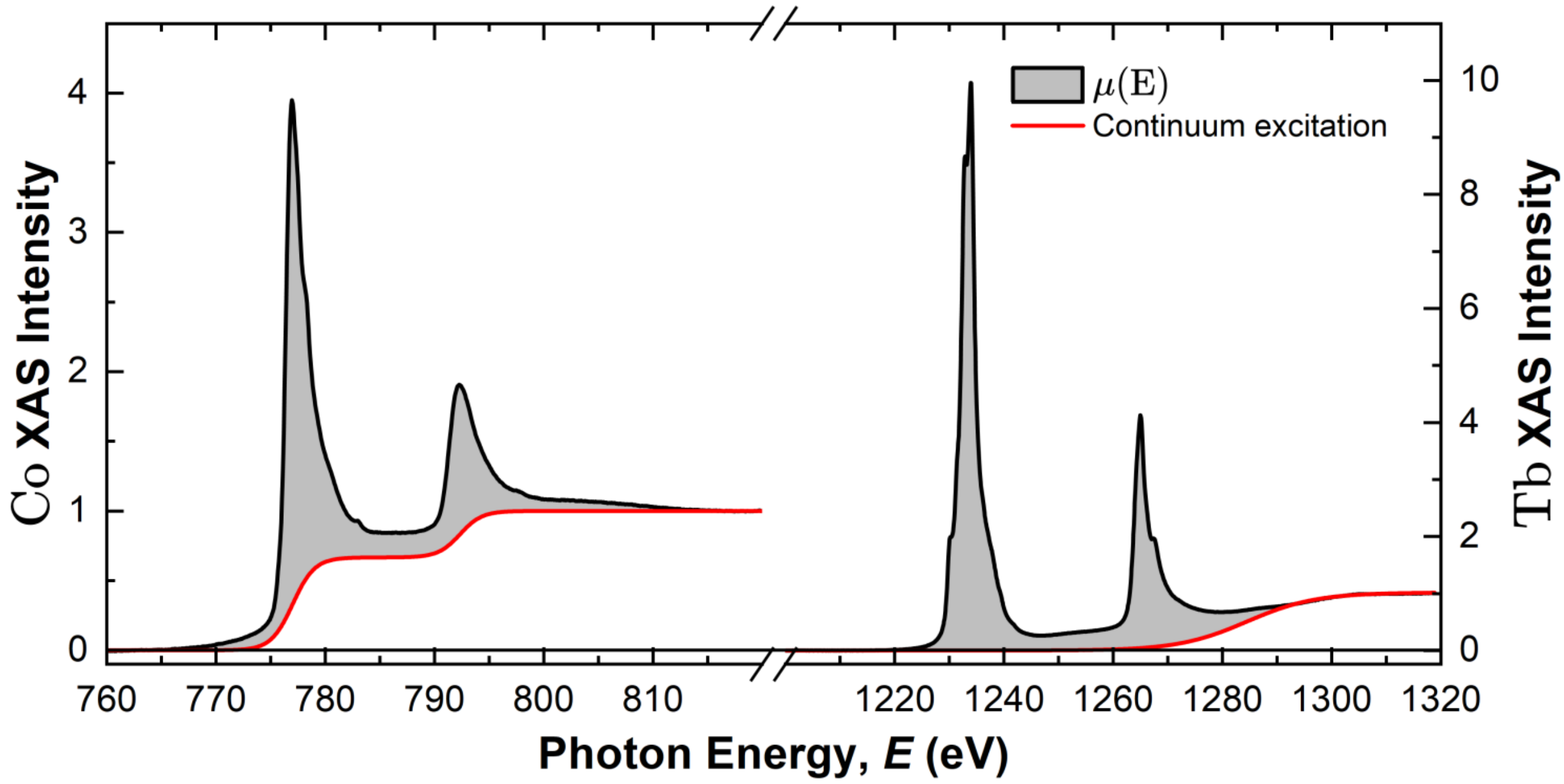

**Figure S11** Step functions for Co (left, red curve) and for Tb (right, red curve) used to account for the transitions into the continuum. The step functions were subtracted from the measured X-ray absorption spectrum (black) to form the integrant $(\mu^+ + \mu^-)$ as the denominator in the formulated sum rules.

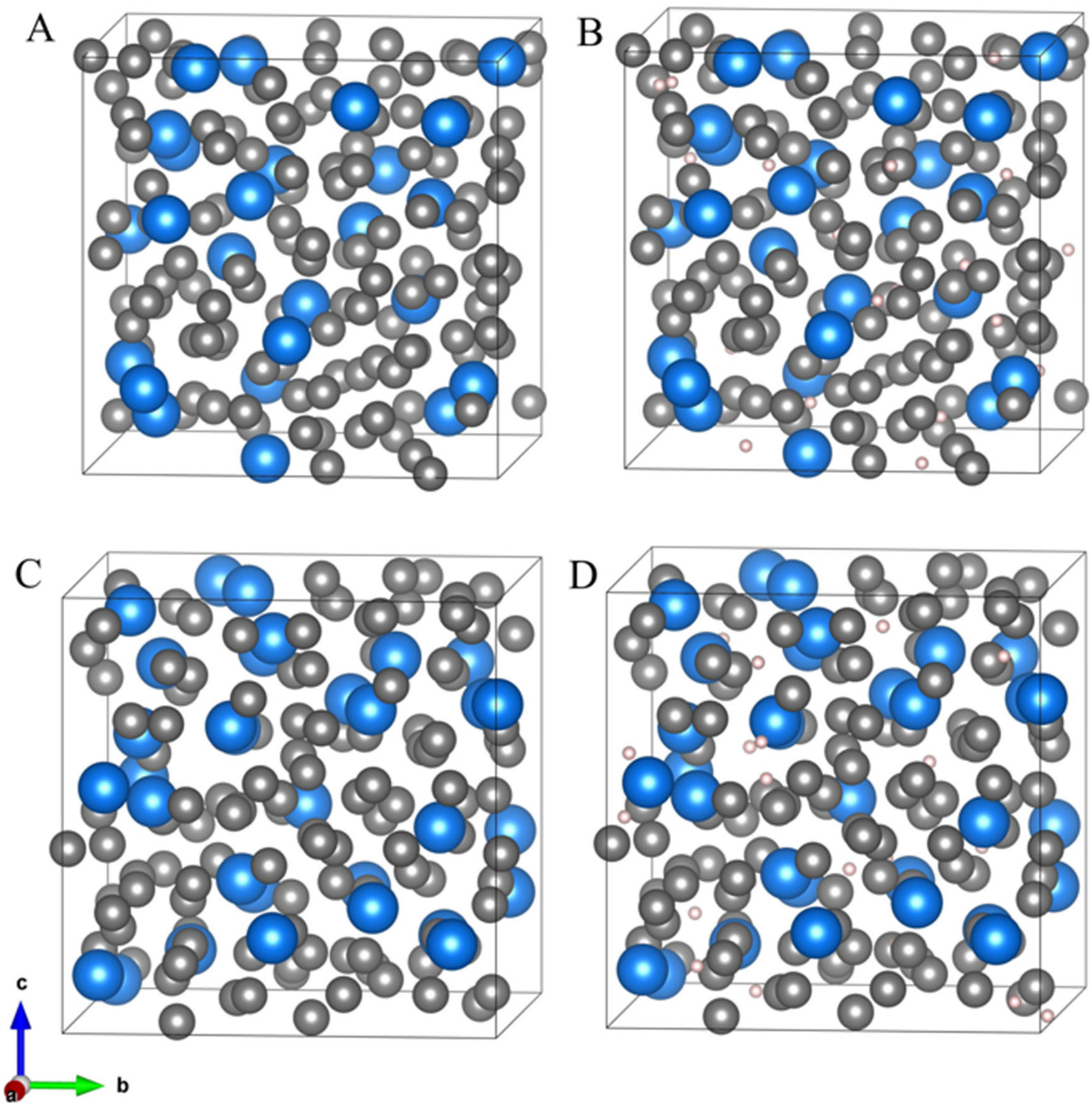

**Fig. S12** The built supercells to simulate amorphous atomic structure of (A) Tb25Co125, (B) Tb25H25Co125, (C) Tb32Co118 and (D) Tb32H32Co118. Blue, grey and pink atoms denote Tb, Co and H atoms, respectively.

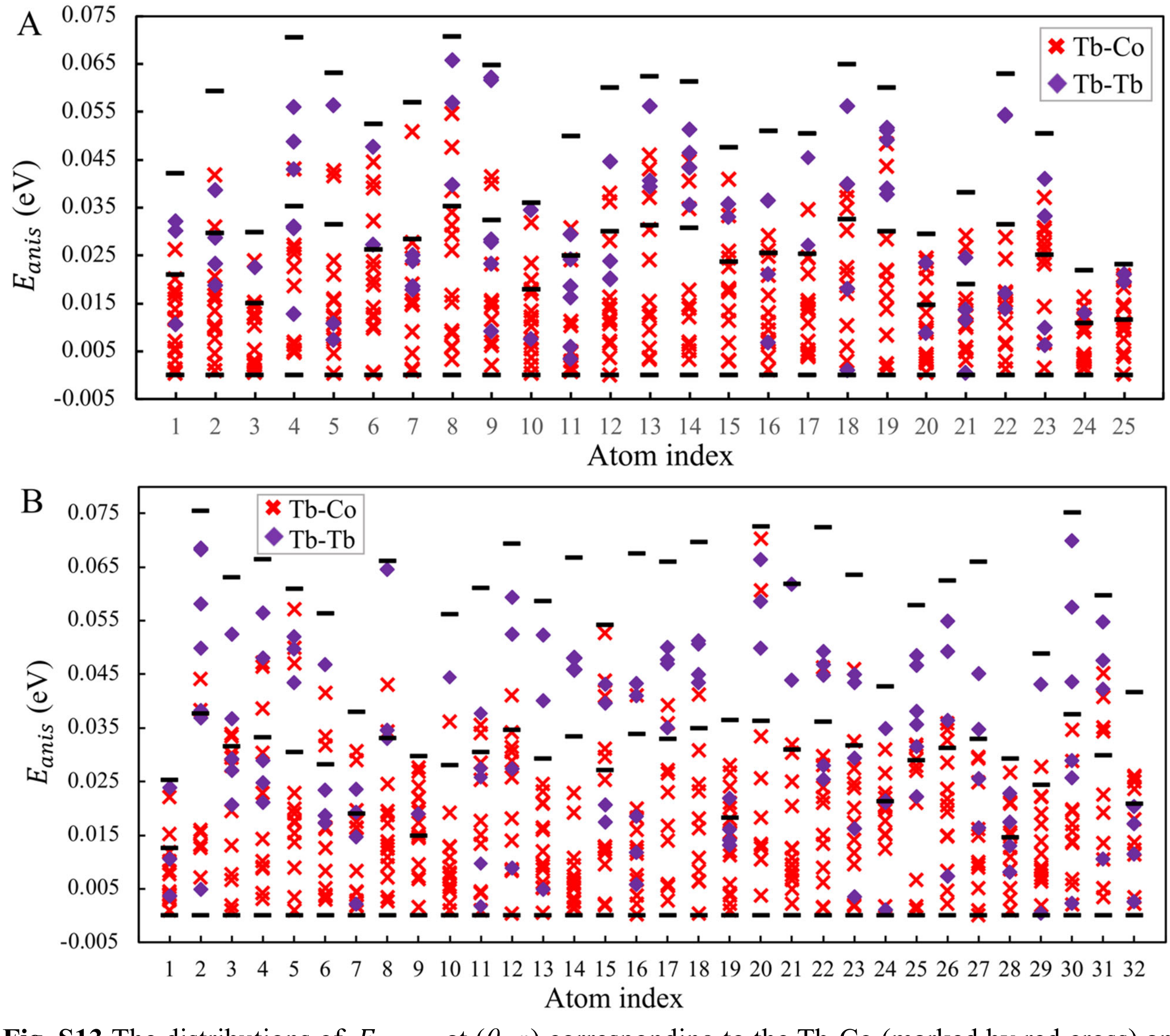

**Fig. S13** The distributions of $E_{aniso}$ at $(\theta, \varphi)$ corresponding to the Tb-Co (marked by red cross) and Tb-Tb (marked by purple diamonds) bonds within the cluster for each Tb atom in (A) Tb25Co125 and (B) Tb32Co118. The minimum, maximum and their averages of $E_{aniso}$ are marked by the black bars. The cutoff radii for Tb-Co and Tb-Tb bonding pairs are 3.5 and 4.0 Å, respectively.

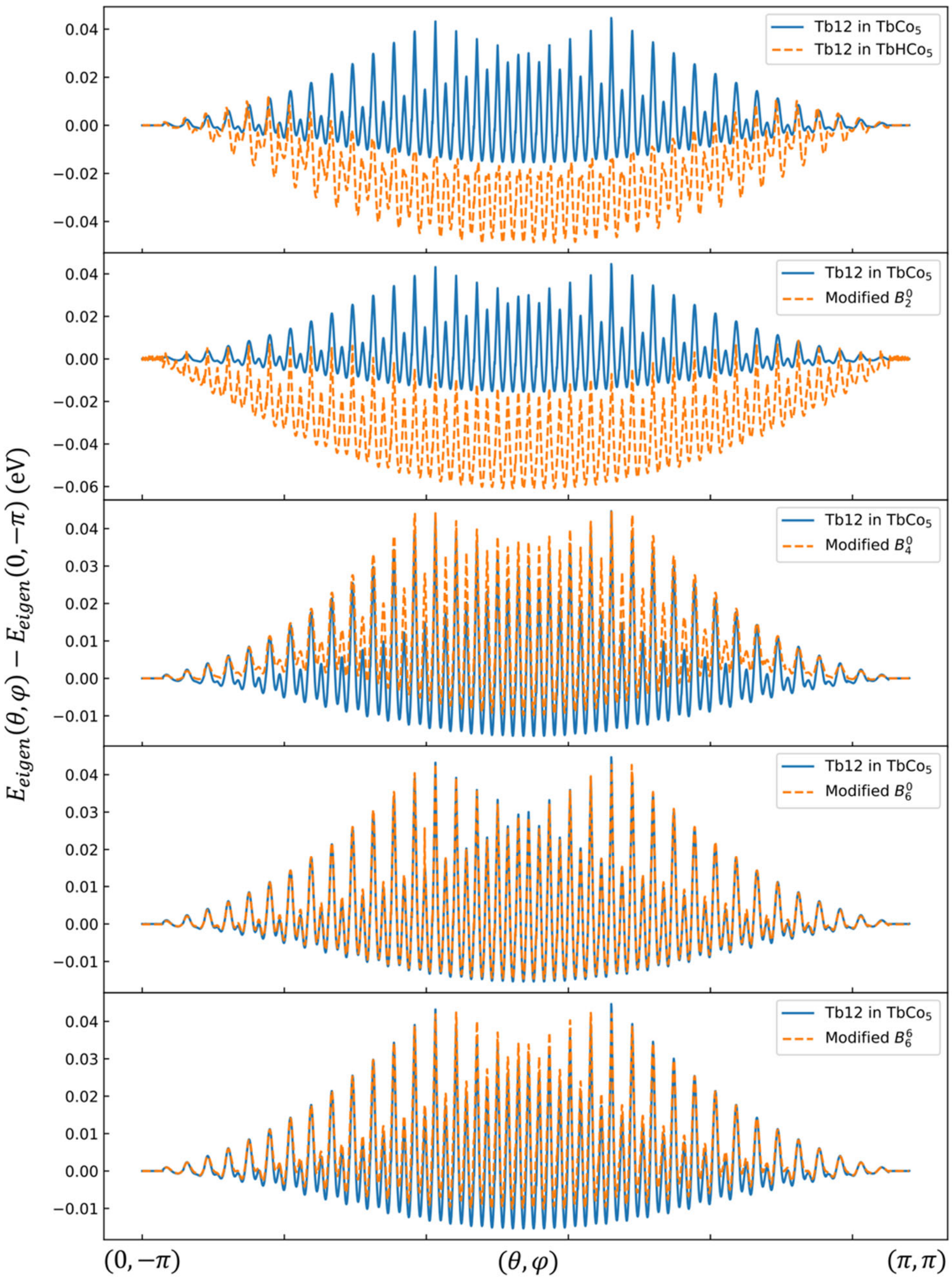

**Fig. S14** $E_{eigen}$ of Tb12 as a function of magnetization direction ($E_{eigen}(0,-\pi)$ as the reference, $E_{eigen}(0,-\pi) = 0$). From upper to lower panel, the comparisons between pristine Tb25Co125 with Tb25H25Co125, Tb25Co125 with modified $B_2^0$, Tb25Co125 with modified $B_4^0$, Tb25Co125 with modified $B_6^0$, and Tb25Co125 with modified $B_6^{\pm 6}$ are shown.

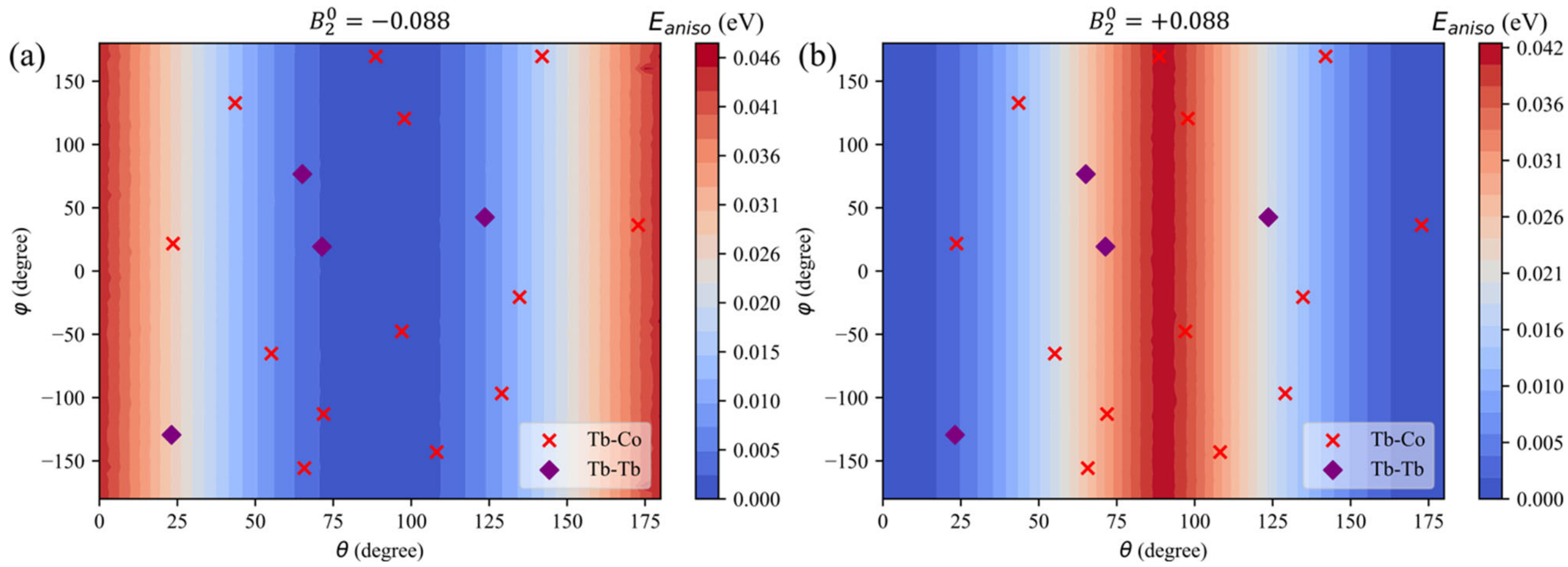

**Fig. S15** Contour of $E_{eigen}$ of Tb12 as a function of magnetization direction ($E_{eigen}(0, -\pi)$ as the reference, $E_{eigen}(0, -\pi) = 0$) with (a) $B_2^0 = -0.088$ and (b) $B_2^0 = +0.088$ while omitting all the other higher-order CFPs.

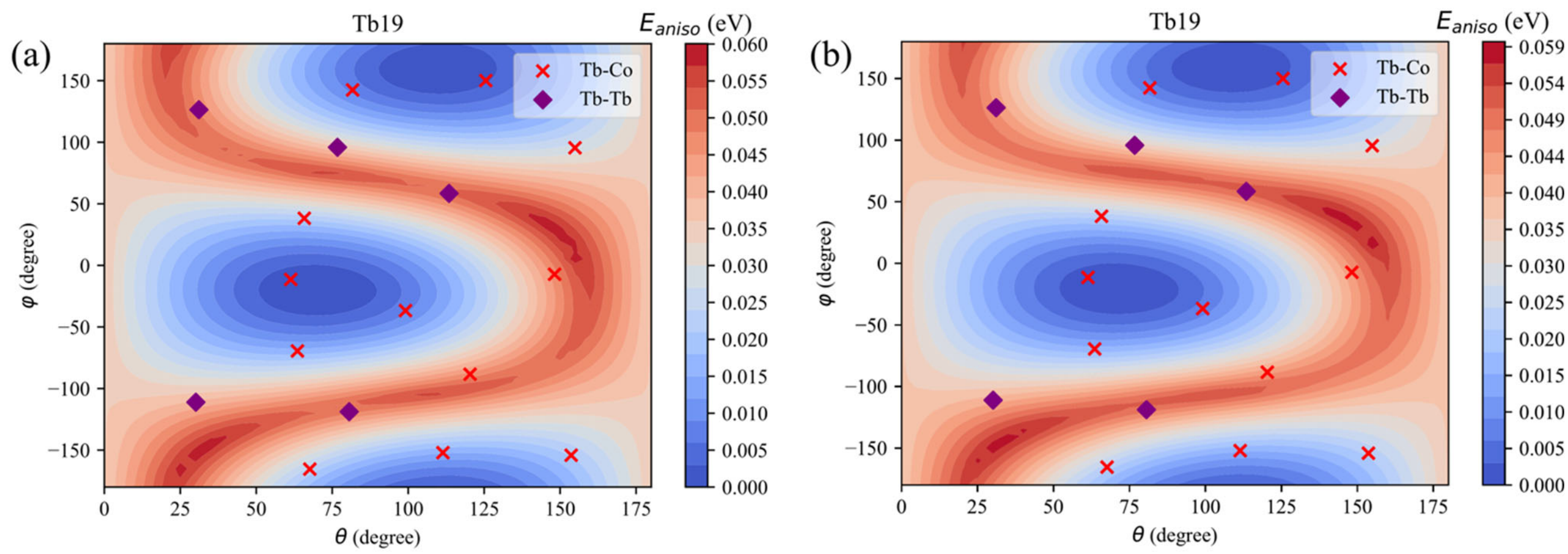

**Fig. S16** Contour maps of $E_{aniso}$ plotted against θ and φ for Tb19 (a) without and (b) with strain. We performed ab inito calculations to investigate the potential strain effect on magnetic anisotropy using an elongated atomic structure along perpendicular direction, which simulates the condition that the in-plane lattice is constraint that leaves only the out-of-plane degree of freedom. The strain level was set to 2% and 5% in the ab initio calculations, respectively. The results show that the magnitudes of crystal-field parameters (CFPs) are altered by around 0.001 eV (less than 5% of the CFPs before structural elongation) under strain up to 5%. It can be seen that the contour map of $E_{aniso}$ with and without strain demonstrate negligible changes due to the strain effect.

**Table S1**

The resolved spin and orbital magnetic moments of Co and Tb in Tb-o films in units of μB/atom, along grazing and perpendicular directions before and after hydrogen insertion. The angles were with respect to the normal of the sample surface.

| | 15° | 15° | 70° | 70° |
|---|---|---|---|---|
| Co | TbCo | TbCoH | TbCo | TbCoH |
| Temperature | 5K | 5K | 5K | 5K |
| $\mu_{spin}(\mu_B/atom)$ | -2.26 | -1.84 | -1.13 | -1.35 |
| $\mu_{orb}(\mu_B/atom)$ | -0.23 | -0.29 | -0.08 | -0.15 |
| | | | | |
| Tb | TbCo | TbCoH | TbCo | TbCoH |
| Temperature | 5K | 5K | 5K | 5K |
| $\mu_{spin}(\mu_B/atom)$ | 5.16 | 3.43 | 3.67 | 4.81 |
| $\mu_{orb}(\mu_B/atom)$ | 3.73 | 3.09 | 3.54 | 3.34 |

**Table S2.** Crystal field parameters (CFPs) of Tb12 in amorphous Tb25Co125 and Tb25H25Co125.

| CFP(eV) | $B_2^0$ | $B_2^{\pm1}$ | $B_2^{\pm2}$ | $B_4^0$ | $B_4^{\pm1}$ | $B_4^{\pm2}$ | $B_4^{\pm3}$ | $B_4^{\pm4}$ |
|---|---|---|---|---|---|---|---|---|
| $TbCo_5$ | 0.022 | 0.026-0.031$i$ | 0.008+0.070$i$ | -0.034 | 0.024-0.016$i$ | 0.001+0.001$i$ | -0.016+0.001$i$ | 0.008-0.002$i$ |
| $TbHCo_5$ | -0.088 | 0.048-0.019$i$ | -0.003+0.043$i$ | -0.148 | -0.005+0.005$i$ | 0.063-0.022$i$ | 0.024-0.007$i$ | 0.099-0.056$i$ |
| | $B_6^0$ | $B_6^{\pm1}$ | $B_6^{\pm2}$ | $B_6^{\pm3}$ | $B_6^{\pm4}$ | $B_6^{\pm5}$ | $B_6^{\pm6}$ | |
| $TbCo_5$ | -0.011 | 0.010+0.013$i$ | 0.026-0.003$i$ | 0.009-0.001$i$ | 0.015+0.005$i$ | 0.017-0.019$i$ | 0.005-0.002$i$ | |
| $TbHCo_5$ | 0.087 | -0.009+0.007$i$ | 0.015+0.001$i$ | -0.009-0.004$i$ | 0.210+0.003$i$ | 0.061-0.004$i$ | 0.186-0.034$i$ | |